\documentclass[10pt]{emulateapj}

\shorttitle{From binaries to multiples II: Statistics}
\shortauthors{Tokovinin}

\begin{document}

\title{From binaries to multiples II: hierarchical multiplicity of F
  and G dwarfs}

\author{Andrei Tokovinin}
\affil{Cerro Tololo Inter-American Observatory, Casilla 603, La Serena, Chile}
\email{atokovinin@ctio.noao.edu}

\begin{abstract}
Statistics  of hierarchical multiplicity  among solar-type  dwarfs are
studied using the distance-limited sample of 4847 targets presented in
the  accompanying Paper~I.  Known  facts about  binaries (multiplicity
fraction  0.46,  log-normal  period  distribution with  median  period
100\,yr and logarithmic dispersion  2.4, and nearly uniform mass-ratio
distribution  independent of  the period)  are confirmed  with  a high
statistical significance.   The fraction of hierarchies  with three or
more  components is  $0.13 \pm  0.01$, the  fractions of  targets with
$n=1,2,3,\ldots$  components  are  54:33:8:4:1.   Sub-systems  in  the
secondary  components  are  almost  as  frequent  as  in  the  primary
components, but  in half of such  cases both inner  pairs are present.
The high frequency  of those 2+2 hierarchies (4\%)  suggests that both
inner  pairs were  formed  by  a common  process.   The statistics  of
hierarchies can be reproduced  by simulations, assuming that the field
is  a mixture  coming from  binary-rich and  binary-poor environments.
Periods of the outer and  inner binaries are selected recursively from
the same log-normal distribution,  subject to the stability constraint
and  accounting for  the correlation  between inner  sub-systems.  The
simulator can  be used to  evaluate the frequency of  multiple systems
with  specified  parameters.   However,  it  does  not  reproduce  the
observed excess  of inner periods  shorter than 10\,d, caused  by tidal
evolution.
%
%
\end{abstract}


\section{Introduction}
\label{sec:intro}

Stars tend  to be born in  pairs and higher-order  hierarchies. It has
long  been  recognized   that  understanding  stellar  multiplicity  is
important  in many  areas of  astronomy,  from star  formation to  
high-energy astrophysics. Here we address the statistics of
hierarchical multiples (triple and quadruple systems) among solar-type
dwarfs, using the data of Paper I \citep{PaperI}.

Hierarchical   stellar  systems   have been   known  since   a  long   time.
\citet{Batten} highlighted  their significance  as a potential  key to
solving  the  mystery   of  binary-star  formation  \citep[see  recent
  discussion in][]{PP6}.  \citet{Fekel81} published the first list of
hierarchical  triples.  Meanwhile,  other authors  who  studied binary
statistics  mentioned  multiples  only  in passing.   Typically  small
samples of $\sim 200$ stars in  these early works contained only a few
hierarchies.

The paper by \citet[][AL76]{AL76} demonstrated the importance of binary
statistics and  influenced the  field for several  decades. By  a bold
extrapolation  it predicted that  almost every  solar-type star  has a
low-mass (planetary)  companion.  This  prediction, shown later  to be
the  result of data  over-interpretation, stimulated  nevertheless the
search for very low-mass (VLM) companions and exo-planets.

The work  by \citet[][hereafter DM91]{DM91} remains even  today one of
the  most  cited papers  on  binary  statistics.   They corrected  the
overly-optimistic prognosis of the  frequency of VLM companions, while
still  focusing  on  their  importance  (few  years  later  this  team
announced the  first exo-planet around  51~Peg). A similar study  of K
and M  dwarfs \citep{Tok92} reached  the opposite conclusion  that the
VLM  companions are  rare; this  is now  confirmed and  known  as {\em
  brown-dwarf  desert}.    The  log-normal  distribution   of  periods
proposed in DM91 was confirmed by further studies, including this one.

Two decades later, \citet[][hereafter R10]{R10} revised the results of
DM91 using  better observing  methods and a  larger sample.  They show
that the  mass ratio is  distributed almost uniformly at  all periods,
overturning  the claims to  the contrary  made in  AL76 and  DM91. The
frequency  of hierarchical  systems  was estimated  at  $f_H =  12$\%,
doubled in comparison to DM91. We show below that it is even a little higher.

Most known multiple stars result from random discoveries.  Information
on  triples and higher-order  systems collected  in the  Multiple Star
Catalog,  MSC  \citep{MSC}   is  heavily  distorted  by  observational
selection   and   not  suitable   for   unbiased  statistical   study.
\citet{Tok08} attempted comparative analysis of triples and quadruples
based   on   the   MSC.    A   similar  effort   was   undertaken   by
\citet{Eggleton09}, using  bright stars and  an {\it ad hoc}  model of
observational selection.

This work is based on  the volume-limited sample of 4847 unevolved (or
moderately evolved)  stars within 67\,pc  of the Sun with  masses from
0.9 to  1.5\,$M_\odot$, presented in  Paper I and called  FG-67 sample
hereafter.  Its  completeness is no less  than 90\%\footnote{The range
  of primary  masses in  the FG-67  sample is less  than in  the 25-pc
  sample, so the number of targets  in those samples does not scale as
  cube of their distance limits.}
As  hierarchical systems are  less frequent  than binaries,  the large
sample size and  a good companion census are  essential here.  Equally
important is  the correction  for the incomplete  companion detection.
Paper~I  outlines  the model  of  companion's  detection and  provides
estimates of detection  probability (completeness) for each individual
target, as a  function of period and mass ratio.   On average, 80\% of
companions to the  main targets are detected.  However,  the census of
sub-systems in the fainter, secondary components is worse, about 30\%.

We begin with the definition  of hierarchical levels and mass ratio in
\S~\ref{sec:h}.   Then in  \S~\ref{sec:pq} the  joint  distribution of
period  and   mass  ratio   is  derived.   In   \S~\ref{sec:trip}  the
statistical  model of  hierarchies is  developed. It  is based  on the
simple  idea that  all sub-systems  are randomly  drawn from  the same
generating  distribution of  periods,  subject only  to the  dynamical
stability constraint.   This recipe requires,  however, some adjustments
to match the data.  The summary and discussion in \S~\ref{sec:sum} and
\S~\ref{sec:disc}, respectively, close the paper.

\section{Description of hierarchical systems}
\label{sec:h}

\subsection{Definition of a hierarchical system}

A  system of  three  or  more bodies  with  comparable separations  is
dynamically   unstable    \citep{Harrington}.    In   contrast,   {\em
  hierarchical} multiples  survive for a long time  because the motion
in  the close  (inner) pair  is not  strongly perturbed  by  the outer
companion(s).  Stellar  motions in stable  hierarchies are represented
approximately  by  Keplerian orbits.   This  is  the  definition of  a
hierarchical  multiple  system.  This  definition  is  not very  sharp
because some  systems approach the stability  boundary.  Moreover, the
orbital motion of very wide  binaries is not observable owing to their
long  periods, so  the  borderline between  the  truly bound  multiple
systems  and  other stellar  groups  like  disintegrating clusters  is
fuzzy.

\subsection{Hierarchical levels and multiplicity fractions}

\begin{figure}
\plotone{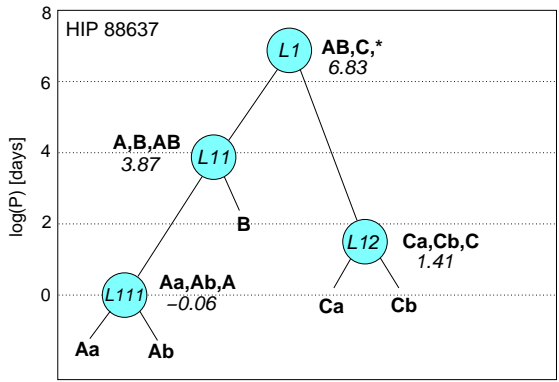}
\caption{Hierarchical  levels  and   companion  designations  for  the
  quintuple system HIP~88637 are represented by a mobile diagram where
  vertical  position  of  each   sub-system  reflects  its  period  $x
  =\log_{10} P/1{\rm  d}$. The hierarchy is coded  by designating each
  system as  {\em (primary, secondary, parent)},  asterisk meaning the
  root.   Levels are  computed automatically  from  such designations.
  The  outermost pair  AB,C,*  ($x =  6.83$)  is at  the  root of  the
  hierarchy, at  L1.  The sub-system  in its primary  component A,B,AB
  ($x=3.87$) is  at L11, the  secondary sub-system Ca,Cb,C is  at L12,
  and the innermost binary Aa,Ab,A is at L111.
\label{fig:levels} }
\end{figure}

Hierarchies can  be described by  binary graphs or trees,  also called
mobile diagrams (one example is shown in Figure~\ref{fig:levels}). The
position of each pair in the tree is called {\em level} and is denoted
here as L1,  L11, etc.  The outermost (widest) pair is  at the root of
the  tree,  L1  (the  root  is  denoted  by  asterisk  in  the  parent
designation).  Inner  pairs associated with the  primary and secondary
components are at levels L11  and L12, respectively, and this notation
continues to deeper levels. Triple  systems can have inner pair either
at L11  or at  L12.  When both  primary and secondary  sub-systems are
present, we get the  so-called {\em 2+2 system}.\footnote{A 2+2 system
  can contain higher hierarchical levels, hence more than 4 stars, see
  Figure~\ref{fig:levels}.   In  contrast,  a 2+2  quadruple  contains
  exactly  4  stars.}  Alternatively,  a  {\em  3+1 quadruple}  system
consists  of  levels  L1,  L11,  and L111;  it  has  three  companions
associated with  the same primary  star, resembling in this  respect a
planetary  system.  Both types  of hierarchy  are found  in HIP~88637.
The assignment of ``primary'' and ``secondary'' is traditionally based
on  the  apparent  brightness   of  the  components.   When  they  are
comparable in brightness or  mass, the distinction between primary and
secondary becomes irrelevant.

While  we deal  with binary  and  triple systems,  their hierarchy  is
uniquely defined  by the  number of components,  so levels  are almost
redundant.  On the other  hand, quadruple and higher-order systems can
have different  organizations, so their  hierarchy requires additional
descriptors such as  levels. From the observational point  of view, it
is easier to  explore lower hierarchical levels L1,  L11, and L12 than
to determine the  number of multiple systems with  {\em exactly} 3, 4,
etc.  components.

The  fraction of  non-single  stellar systems,  the {\em  multiplicity
  fraction}  $f_M$, equals  the  fraction  of L1  systems  in a  given
sample.  The {\em  companion fraction} $f_C$ is the  number of systems
at all levels divided by the sample size, so $f_C > f_M$. It is easier
to measure $f_M$, as it does  not depend on the discovery of {\em all}
sub-systems.  Multiplicity  is also characterized by  the fractions of
systems with exactly $n$ components, $f_n$. Obviously, the fraction of
single stars  $f_1 = 1  - f_M$,  while $f_C =  f_2 + 2  f_3 + 3  f_4 +
\dots$  The   fraction  of   hierarchies  in  a   sample  is   $f_H  =
\sum_{n=3}^{\infty} f_n = 1 -  f_1 - f_2$. The multiplicity statistics
are defined by  $f_n$, by the distribution of levels,  and by the joint
distributions of periods and mass ratios at different levels.

The system  depicted in Figure~\ref{fig:levels}  contains hierarchical
levels most frequently  found in the FG-67 sample:  L11 ($N=296$), L12
($N=95$), and  L111 ($N=17$).  Here the statistical  analysis of those
three levels is made, because higher levels are rare, and a much larger
sample would be needed for their study.

\subsection{Mass ratio definition}
\label{sec:qdef}

In  a binary system  with the  primary component  A and  the secondary
component B,  the mass  ratio is defined  as $q  = M_B/M_A <1$.   In a
triple  system where B  is a  close pair  B,C, the  mass ratio  can be
defined  in  different   ways  \citep[e.g.][]{Tok08}.   The  classical
approach is to use the mass sum of sub-systems in the {\em system mass
  ratio} $q_{\rm  sys} = (M_B  + M_C)/M_A$.  However,  this definition
may result  in $q_{\rm sys}>1$ even  when the component A  is the most
massive  and brightest star  in the  system.  Alternatively,  the {\em
  component  mass ratio}  $q =  M_B/M_A$ takes  into account  only the
masses   of  individual  stars   (primary  components   of  respective
sub-systems). For example, in a  triple star consisting of the close
binary A,B with  a distant tertiary component C,  $q = M_C/M_A$, while
$q_{\rm  sys} =  M_C/(M_A +  M_B)$.  The  detection limits  of imaging
techniques are based on the magnitude difference which is more tightly
related to  the component mass ratio  $q$ than to  $q_{\rm sys}$.  The
use of $q$ avoids  re-definition of ``primary'' and ``secondary'' that
would be needed in many instances if we used the mass sum instead.  In
the following,  we use consistently  the component mass ratio  $q$ for
multiple systems.

\subsection{Statistical data on the FG-67 sample}

\begin{figure}
\plotone{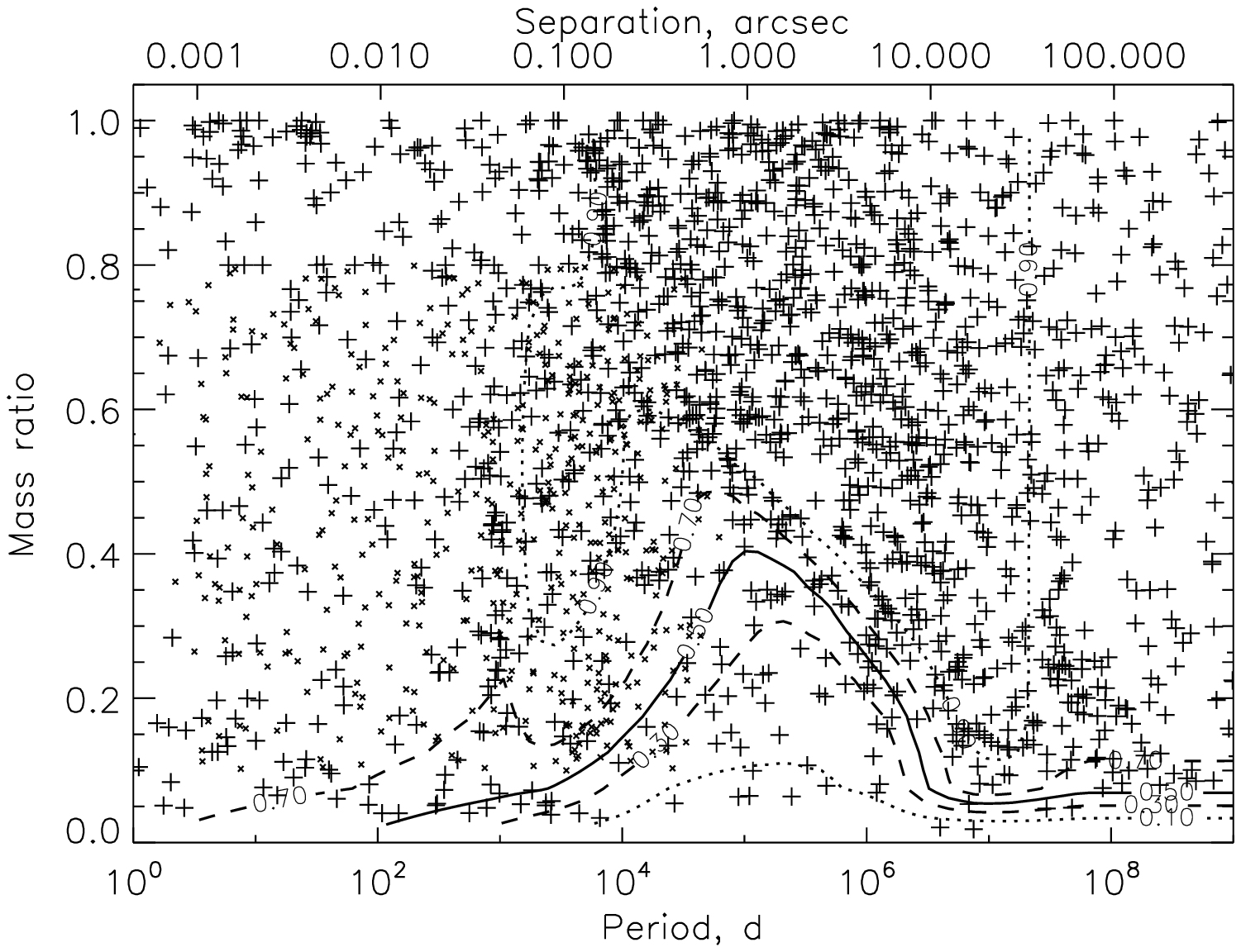}
\plotone{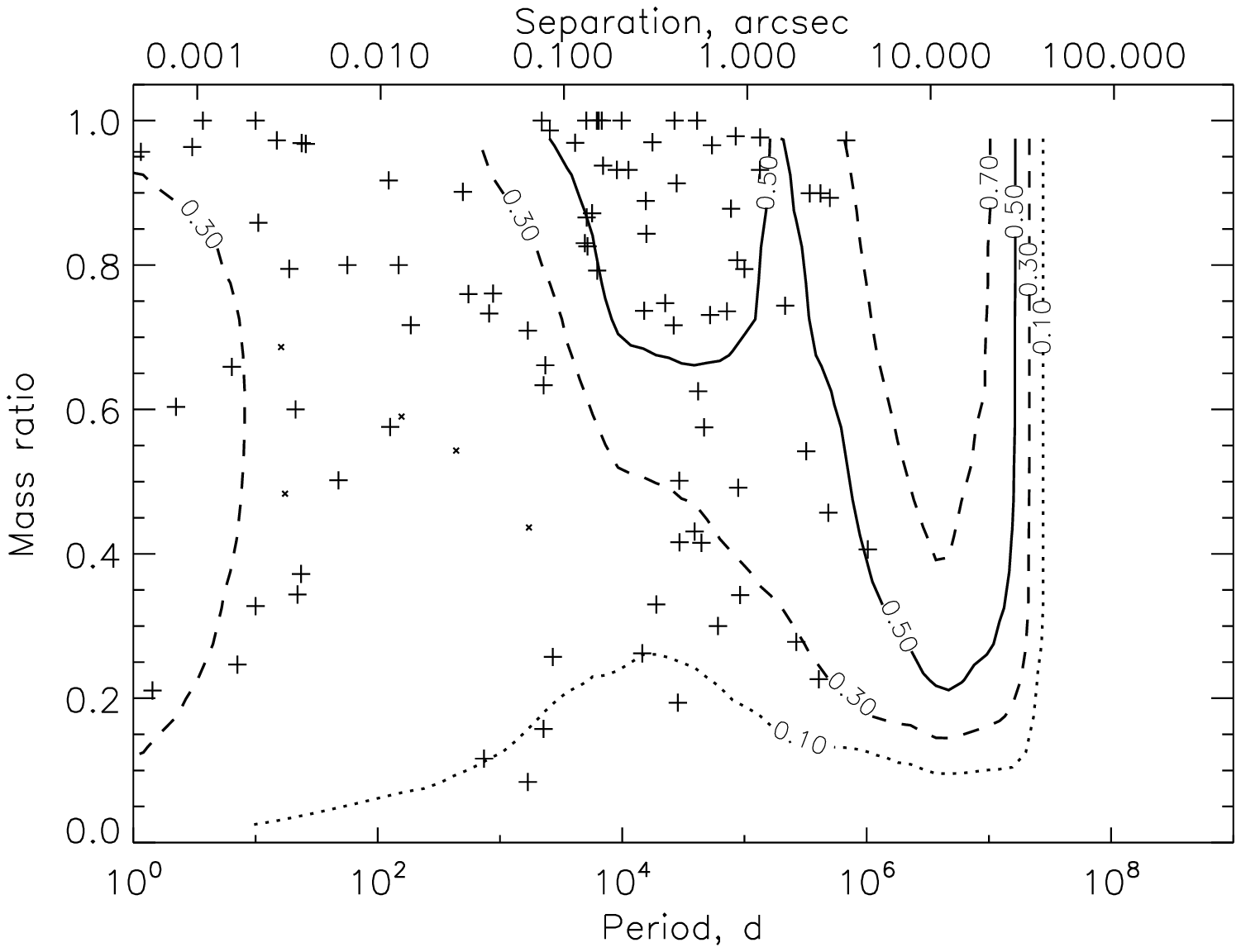}
\caption{Distribution  of systems around  primary targets  (top panel)
  and sub-systems  in the secondary  components (bottom panel)  in the
  $(P,q)$  plane.   Plus  signs  --  known  $P,q$,  small  crosses  --
  fictitious values.  The contours indicate completeness of
  0.1, 0.3, 0.5 (full line),  0.7, and 0.9. The upper axis corresponds
  to the angular separation at a distance of 50\,pc.
\label{fig:pq} }
\end{figure}

The  data on  all  systems and  sub-systems  in the  FG-67 sample  are
assembled in  the structure (table) which contains  information on the
hierarchical level  (with links to parents and  siblings), period, and
component's masses.  The same  structure is produced for the simulated
samples.  As binary periods $P$ span  a huge range, we use instead the
logarithm $x = \log_{10} (P/1d)$ throughout this paper.

The 22 targets containing white dwarfs (WDs) are removed, leaving 2162
systems  and  sub-systems to  consider  (the  sample excludes  evolved
primary  components by design,  WDs are  such former  primaries).  The
fraction of targets  with WD companions is from  2\% to 4\%, according
to the rather  uncertain estimate of Paper~I.  As  only 0.5\% of known
binaries are removed  and most of the remaining  WDs are undiscovered,
the total sample size without WDs should be somewhere between 4650 and
4750.  The sample  size affects  the estimates  of $f_M$,  $f_C$, etc.
Here we  set it  conservatively to 4800  (i.e.  remove 1\%  of targets
with WDs).  The multiplicity fraction  derived here would be larger by
$\sim$1\% if more targets with WDs were removed.  Interestingly, among
the  22 binaries  with  WDs, 11  are  at least  triple, their  FG-type
primaries (former secondaries) contain sub-systems.  This supports the
large fraction of secondary sub-systems found here in the main sample.

\begin{deluxetable}{r  l l c  c c c c} 
\tabletypesize{\scriptsize}      
\tablecaption{Statistical data (fragment).\tablenotemark{a}
\label{tab:stat}   }     
\tablewidth{0pt} 
\tablehead{ HIP0 & L & Type\tablenotemark{b} & $x$ &  $M_1$ &  $M_2$ & $M_{1,p}$ & $M_{2,p}$ \\
                &       &      & d   & $M_\odot$ &   $M_\odot$ &     $M_\odot$ & $M_\odot$ }
\startdata
223   &1   &v    &  5.16&  1.17 & 0.84&  1.17&  0.84\\
290   &1   &s,a  &$-$9.00& 1.28 & 0.00&  1.28&  0.00\\
359   &1   &a    &$-$9.00& 0.96 & 0.00&  0.96&  0.00\\
394   &1   &S1   &  2.84&  1.43 & 0.00&  1.43&  0.00\\
493   &1   &Cpm  &  8.83&  1.09 & 1.67&  1.09&  0.88\\
493   &12  &Ch   &  5.62&  0.88 & 0.79&  0.88&  0.79\\
518   &1   &V    &  4.59&  1.06 & 1.37&  1.06&  0.91\\
518   &12  &S1   &  1.68&  0.91 & 0.46&  0.91&  0.46\\
522   &1   &Chp  &  5.37&  2.25 & 0.54&  1.20&  0.54\\
522   &11  &v    &  3.87&  1.20 & 1.05&  1.20&  1.05
\enddata                                                                
\tablenotetext{a}{Table 1 is published in its  entirety in the electronic
  edition of AJ, a portion is shown here for guidance regarding its
  form and content.}
\tablenotetext{b}{Type means discovery techniques, e.g. s,S1,S2 --
  spectroscopic binaries, a,A -- astrometric binaries, v,V -- visual
  binaries, C -- wide common proper motion pairs; see \S3 of Paper I}
\end{deluxetable}

The  statistical  data  are  presented in  Table~\ref{tab:stat}.   Its
columns contain  the HIP), the  {\em Hipparcos} number of  the primary
component,  hierarchical level L,  type of  the system  (see Paper~I),
logarithm of  the period $x$ (in  days), system masses  of the primary
and  secondary components  $M_1$ and  $M_2$ (in  solar mass),  and the
component  masses  $M_{1,p}$ and  $M_{2,p}$  (primaries  in the  inner
sub-systems).  The two definitions  of mass ratio in \S~\ref{sec:qdef}
correspond  to $q_{\rm  sys}  = M_2/M_1$  and  $q =  M_{2,p}/M_{1,p}$.
Masses are  taken from  the SYS table  of Paper~I, they  are estimated
from  absolute magnitudes  using standard  relation  for main-sequence
stars.  Unknown masses have  zero values, unknown periods are assigned
$x=-9$.  The links to children  and parent (record numbers in the data
structure,  or  $-1$  if  absent) are  generated  automatically  using
component designations; they are not printed in Table~\ref{tab:stat}.

A significant  number, 355 out  of 2162 systems, have  unknown periods
and separations.  These are spectroscopic binaries without known orbit
and  binaries   discovered  by  {\it   Hipparcos}  accelerations  (see
Paper~I).  Masses of their secondary components are also unknown.  The
statistical   analysis  presented  below   deals  with   this  missing
information (32\% of all  binaries with periods shorter than 100\,yr).
The detection probability of these binaries is properly modeled, so we
cannot simply  ignore them.  The  assumption is made that  all unknown
periods  are  shorter  than  $\sim$100\,yr.  Random  {\em  fictitious}
periods  are generated to  replace missing  data when  plotting period
distributions, but  these fictitious  data are not  used in  any other
way.  Considering the detection limits of spectroscopy and astrometry,
fictitious periods of spectroscopic binaries are uniformly distributed
between $0.5  < x <  3.2$, acceleration binaries with  constant radial
velocity  have   $3.2  <  x   <4.5$,  e.g,  HIP~290  and   HIP~359  in
Table~\ref{tab:stat}.  Partially  missing information is  the weakness
of the  FG-67 sample  that affects the  statistics at  periods shorter
than 100\,yr.

\begin{figure}
\plotone{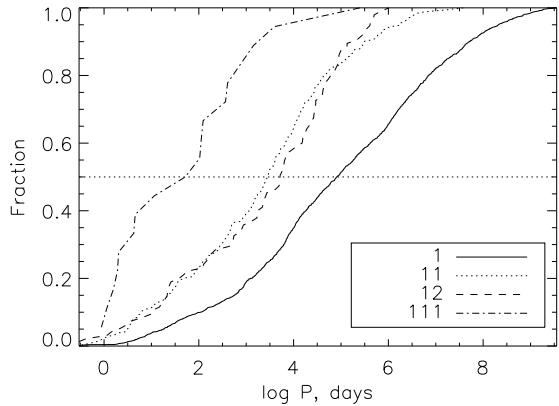}
\caption{Cumulative  histogram  of  periods  at  various  hierarchical
  levels. Unknown periods are complemented by fictitious values. 
\label{fig:pcum} }
\end{figure}

Figure~\ref{fig:pq}  shows  position of  known  companions to  primary
targets and  to the  secondaries in the  $(P,q)$ plane.  Binaries with
unknown  parameters are  plotted by  small symbols  to  indicate their
existence  and likely  location in  this diagram.   The  contours show
average detection probability. The  census of low-mass companions with
separations from 0\farcs3 to  10$''$ is still rather incomplete, other
regions of  the parameter space  are covered better.  The  mass ratios
are  distributed   almost  uniformly  at  all   periods.   The  period
distribution has  a broad  maximum and declines  on both sides  of it.
Note  the  cluster  of  twin  binaries with  $q  \approx  1$  at
$P<100$\,d \citep{twins,Lucy06}.
Figure~\ref{fig:pcum}  shows cumulative  distributions  of periods  at
three hierarchical levels (with fictitious periods included).

\section{Distribution of period and mass ratio}
\label{sec:pq}

At any  given period, the completeness of  companion detection depends
on the  mass ratio.  Therefore,  the distributions of period  and mass
ratio must  be determined jointly,  taking into account  the detection
probability $d(x,q)$ for each target. Integrals of these distributions
give the multiplicity fraction and $f_n$.

Our goal -- the unbiased statistics of hierarchies -- is approached by
two  complementary  methods.   First,  we model  the  distribution  of
binaries  in period and  mass ratio  $f(x,q)$ at  various hierarchical
levels by  analytical functions and  estimate the parameters  of these
models,  accounting  for  the   selection    and  the missing  data
\citep{Tok06,Allen07}.   Second, a statistical  model is  developed by
simulating  the population  of binary  and multiple  stars, simulating
their  incomplete detection, and  comparing the  result with  the real
sample.  In this second approach, adopted e.g.  by \citet{Eggleton09},
the  model has to  be elaborated  by trial  and error.   The simulated
sample represents then the selection-corrected statistics.  The results
of both methods are in mutual agreement.

\begin{figure}
\plotone{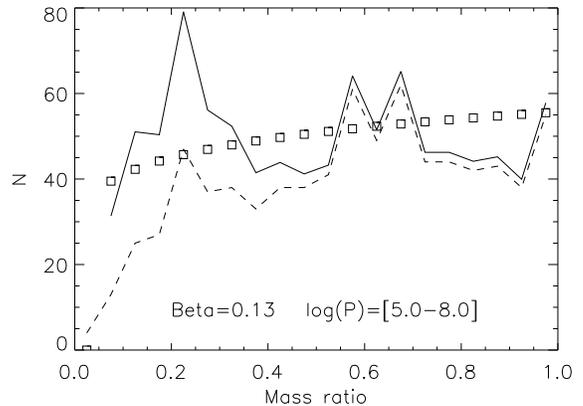}
\caption{Histogram of mass  ratio for 766 wide binaries with  $5 < x <
  8$. Dashed line:  observed, full line: selection-corrected, squares:
  power law.
\label{fig:beta5-8} }
\end{figure}

\subsection{Analytical distribution of $P$, $q$}

Since \citet{DM91} fitted a log-normal function to the distribution of
periods of  82 binaries, following the earlier  work by \citet{Kuiper},
this analytical  model has been universally  adopted. The distribution
of the mass ratio is often approximated by the power law $f(q) \propto
q^\beta$ \citep{DK13}.  It is still being debated whether the exponent
$\beta$  depends on  period  and mass  or  is universal  \citep{RM13}.
Following  the  tide, we  represent  the  period  distribution by  the
truncated log-normal  law and  the $q$ distribution  by the  power law
with  $\beta   >  -1$,  truncated  at   $q<0.05$  (ignore  brown-dwarf
companions).  The analytical model is
\begin{equation}
f(x,q) = C \; \epsilon \; q^\beta \exp [ - (x -
    x_0)^2/(2 \sigma^2)] ,
\label{eq:Gauss}
\end{equation}
where $ x = \log_{10} (P/1d)$. The period distribution is truncated at
$x<-0.3$ and $x>10$ and re-normalized accordingly by the constant $C$;
without truncation,  $C =  (1 + \beta)/(\sigma  \sqrt{ 2  \pi})$.  The
fitted parameters are  multiplicity fraction $\epsilon$, median period
$x_0$,  period  dispersion   $\sigma$,  and  the  mass-ratio  exponent
$\beta$. The assumption that $\beta$  is same at all periods is tested
in the  following sub-section.  The  correspondence between $\epsilon$
and  $f_M$   or  $f_C$  depends   on  the  choice  of   binaries,  see
\S\ref{sec:param}.

The parameters of the model (\ref{eq:Gauss}) were fitted to various
sub-samples using maximum likelihood (ML) method. It takes into account the
detection limits and deals with the missing data. The algorithm and
its implementation are described  in Appendix~\ref{sec:ML}. 

\subsection{Dependence of mass ratio on period?}
\label{sec:beta-p}

\begin{figure}
\plotone{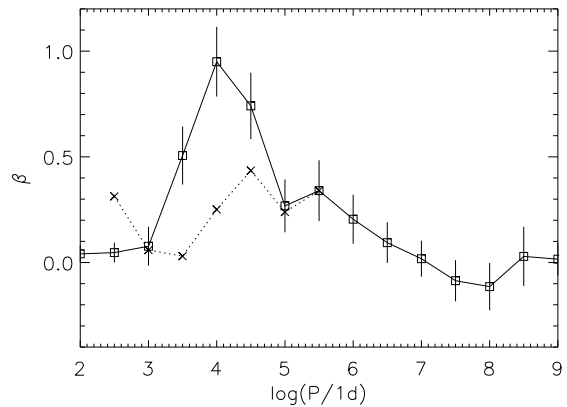}
\caption{Dependence  of  the mass-ratio  exponent  $\beta$ on  period.
  Each point  corresponds to 1\,dex  interval of period.   The crosses
  and  dotted  line  show  alternative  calculation  by  complementing
  unknown periods and mass ratios with their plausible values.
\label{fig:beta} }
\end{figure}

Potential dependence  of mass  ratio on binary  period can  be studied
here, taking  advantage of the  large sample size and  known detection
completeness.  Binaries in certain  period intervals were selected and
fitted with just  two parameters $(\epsilon, \beta)$ by  the ML method
(Appendix  A). The  detection probability  $d(x,q)$ was  averaged over
these period  intervals, leaving only  its dependence on $q$.   We did
not fit  the histograms of $q$,  avoiding any binning,  but they match
the derived $\beta$ reasonably well (Figure~\ref{fig:beta5-8}).

The results  of such calculations in the  overlapping period intervals
of  one decade  are  plotted in  Figure~\ref{fig:beta}. Each  interval
contains from 70 to 300  binaries. Only companions to the main targets
are used  here, regardless of their hierarchical  level. The secondary
sub-systems have different, smaller  completeness, and are too few (95
total). At both  short and long periods, the  distribution of the mass
ratio is uniform, with $\beta  \approx 0$. This matches the results of
other studies \citep{ANDICAM,RM13}.

In the period interval $x =  [3,5]$, the derived $\beta > 0.5$ implies
preference  of large mass  ratios.  Indeed,  most known  binaries with
such periods have $q > 0.5$ because companions with $q<0.5$ have a low
detection   probability  (see  Figure~\ref{fig:pq}).    The  detection
probability  is taken  into  account in  the  calculation of  $\beta$;
however, it ignores binary systems with unknown $P,q$, while including
their discovery  by acceleration and radial velocity  in the evaluated
completeness.  To  access the importance  of this effect,  we assigned
fictitious periods and  mass ratios to the binaries with  missing data and
repeated the  calculation.  The  fictitious mass ratios  are uniformly
distributed in the interval  [0.2-0.8] at periods shorter than $x=3.5$
and in  the interval  [0.2,0.6] at periods  $3.5 <  x < 4.5$  (if such
binaries had larger  mass ratios, they would have  been resolved).  By
complementing the missing data  in this reasonable, but arbitrary way,
we obtain smaller values of $\beta$ in the problematic period interval
$3  < x <  5$ (dashed  line and  crosses in  Figure~\ref{fig:pq}).  We
conclude  that the  large values  of  $\beta$ likely  result from  the
missing data and are not real.
 
There remains a hint on a mild dependence of $\beta$ on period even at
$x>5$, where  there are  no missing data.   However, the range  of this
variation, if any, is small. We fit the parameters of the 
model (\ref{eq:Gauss}) by assuming  a constant $\beta$.  This helps to
minimize   the   effect  of   missing   information  at   intermediate
periods. The  global fitting in \S~\ref{sec:param}  confirms the small
values of $\beta$.

\subsection{Parameters of the distributions}
\label{sec:param}

Table~\ref{tab:par} lists the parameters of the distributions found by
the ML method, for all  pairs without distinction of their hierarchy and
for the sub-samples.   Its second column gives the  number of binaries
selected.   The   errors  corresponding  to   68\%  confidence  intervals
(``1$\sigma$'') are evaluated by  the ML method. We checked  that there is
no strong correlation between  parameters. However, one should keep in
mind that these  errors do not account for any problems  in the input data
or their  modeling.  The formal  ML errors give  a lower bound  of the
true errors.

\begin{table}
\caption{Fitted distribution parameters}
\label{tab:par}
\begin{tabular}{l  c c c c c } 
\hline
\hline
Case & $N$ &  $\epsilon$ & $x_0$ & $\sigma$ & $\beta$  \\
\hline
All pairs & 2162 & 0.571    & 4.54    & 2.40    & 0.094  \\
          &     &$\pm$0.012 &$\pm$0.06&$\pm$0.06&$\pm$0.020  \\
L1   & 1747 & 0.464    & 4.93    & 2.34    & 0.051  \\
          &     &$\pm$0.011 &$\pm$0.06& $\pm$0.06&$\pm$0.011  \\
L11  &  296 & 0.214    & 3.25    & 1.80    & 0.121  \\
          &     &$\pm$0.012 &$\pm$0.12&$\pm$0.09&$\pm$0.026  \\
L12  &   95 & 0.157    & 2.67    & 1.68    & 1.32   \\
          &     &$\pm$0.016 &$\pm$0.17&$\pm$0.10&$\pm$0.28   \\
\hline
\end{tabular}
\end{table}

When all pairs are selected, their fraction $\epsilon=0.57$ equals the
companion  fraction $f_C$.  However,  the code  does not  consider the
reduced  probability   of  detecting  sub-systems   in  the  secondary
components,  hence   delivers  the  slightly   under-estimated  $f_C$;
$f_C=0.65$   follows   from  the   simulations   presented  below   in
\S~\ref{sec:sim}.   When only  the  pairs at  L1  (outer systems)  are
selected, we obtain  the multiplicity fraction $\epsilon =  f_M = 0.46
\pm 0.01$.  It agrees well with  $f_M = 0.46$ found by \citet{R10} for
solar-type stars.

For the sub-systems  at levels L11 and L12, the  presence of the outer
pair  at  L1  is  a  necessary condition.   Therefore,  the  resulting
frequency $\epsilon$  refers to the sample  of L1 systems,  not to the
whole  sample.  The  ML results  indicate that  $0.464 \times  0.214 =
0.100$ fraction  of all  targets have at  least two  companions around
their primary stars. Similarly,  $0.464 \times 0.157 = 0.073$ fraction
of targets have binary secondaries.  This last estimate depends on the
larger  (and  less certain)  correction  for  the  low probability  of
detecting secondary sub-systems.

The total fraction of hierarchical  systems $f_H$ is less than the sum
of  the above  numbers (17\%)  because those  two groups  overlap (see
below).  About 0.6 fraction of L12 (secondary) sub-systems also have a
L11  (primary) sub-system.   The fraction  of hierarchical  systems is
therefore estimated by ML as $f_H  = 0.100 + 0.4\times 0.073 = 0.129$.
A ballpark  estimate of  the uncertainty of  this number  is $\pm$1\%.
Simulations presented in \S~\ref{sec:sim} confirm the derived fraction
of hierarchies. Our initial analysis  (before removing the WDs) gave $f_H =
0.14$.

\begin{figure}
\plotone{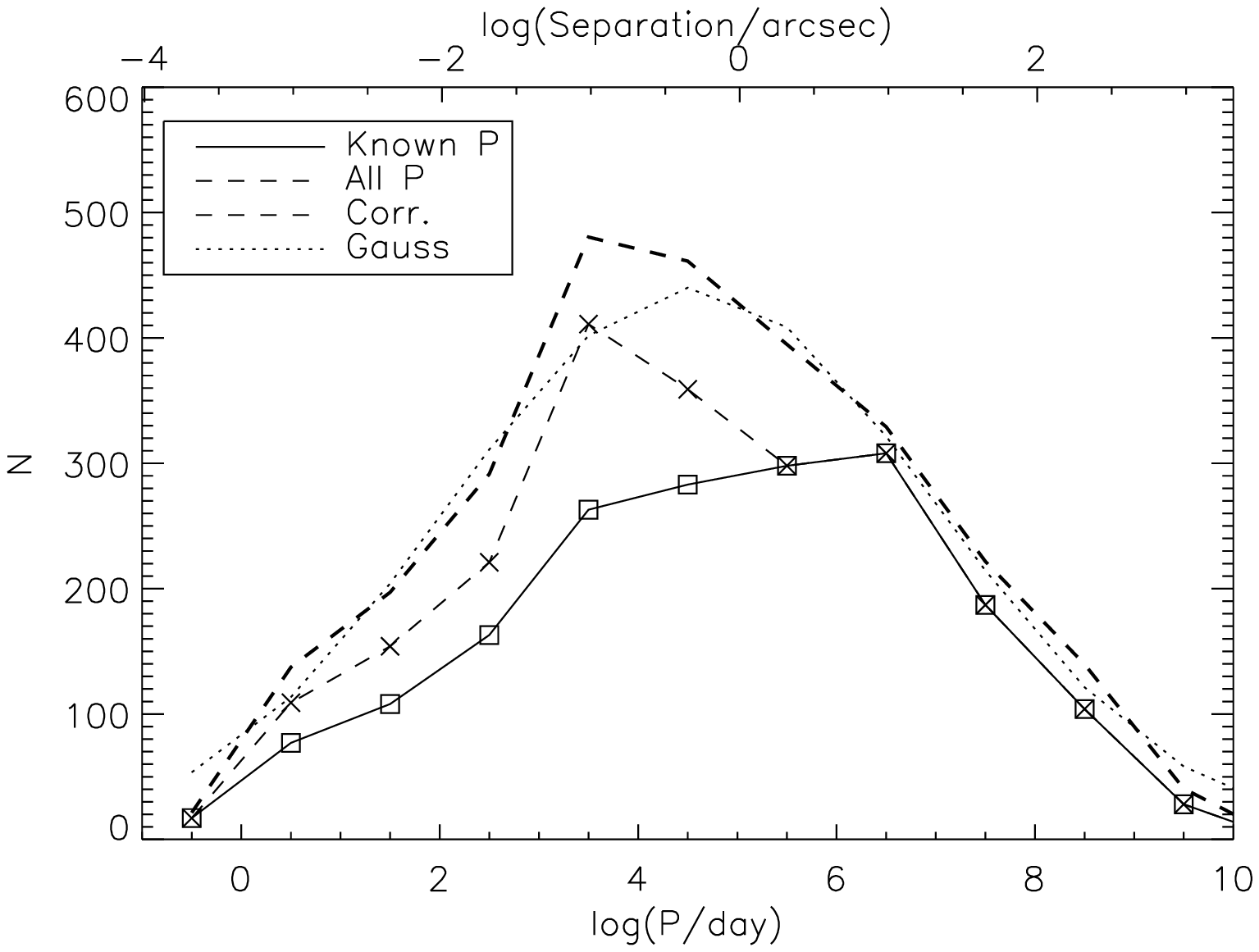}
\plotone{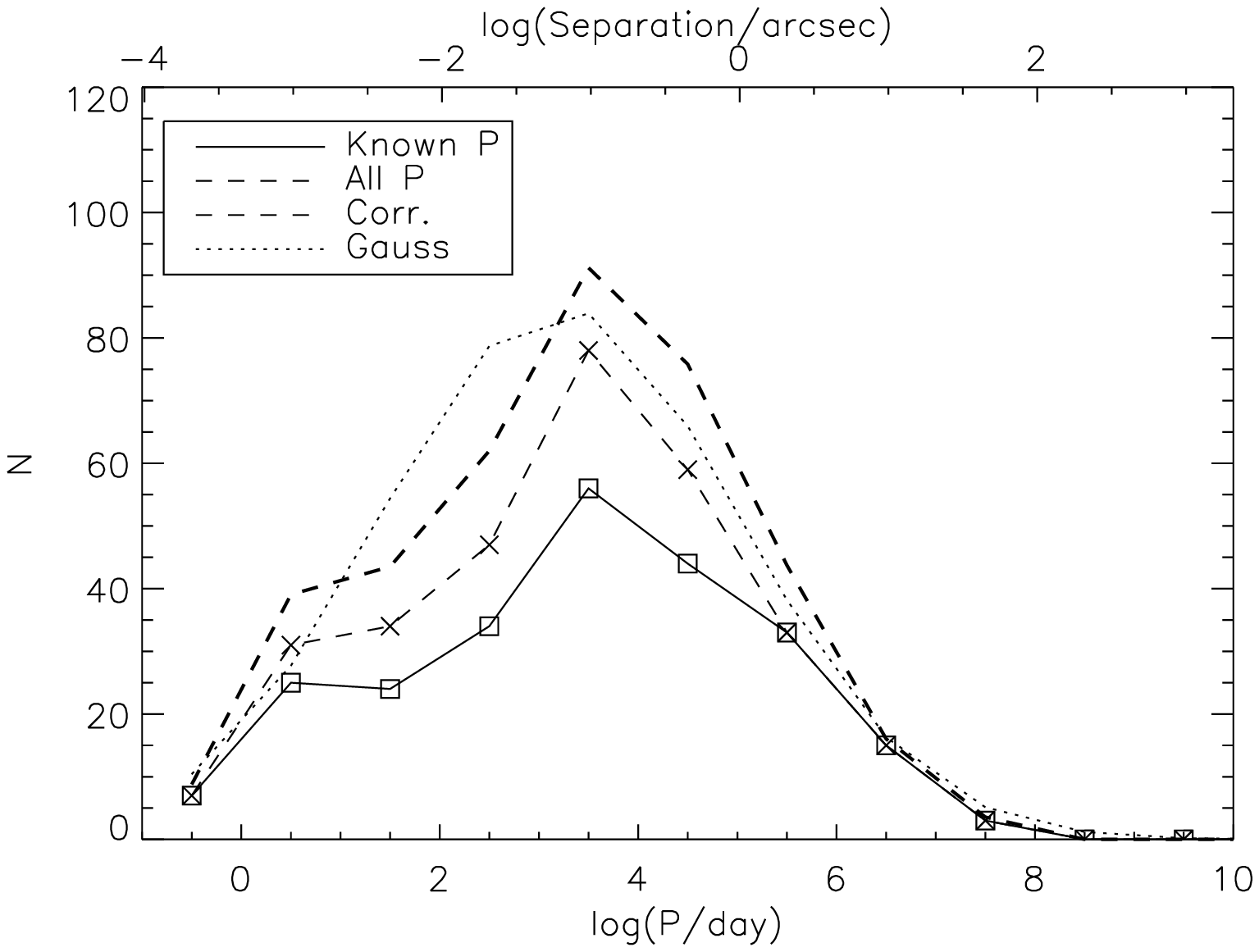}
\caption{Histograms of periods in $\Delta x = 1$ bins. Top: all pairs,
  bottom:  level L11  only.  In  both  plots, the  full  line  is  the
  distribution  of pairs  with known  periods, the  dashed  lines adds
  fictitious periods, the thick dashed line corrects this distribution
  for selection, and the dotted line is the Gaussian distribution with
  parameters  found by  ML, normalized  to match  the total  number of
  systems. Angular separation at a  distance of 50\,pc is indicated on
  the upper horizontal axis of both plots.
\label{fig:phist} }
\end{figure}

Figure~\ref{fig:phist} shows the period histograms of all pairs and of
primary sub-systems  at L11.  Randomly assigned  fictitious periods do
not allow  a rigorous comparison with  the model in  the $x<4.5$ zone.
The  incompleteness  in each  period  bin  is  corrected by  averaging
$d(x,q)$  over  $q$,  which   is  valid  for  the  uniform  mass-ratio
distribution only.  Despite  these shortcomings, the log-normal period
distribution appears to be a good match to the data (the dotted curves
were {\em not} fitted to the histograms!).

Previously, when  the samples of  binaries were small,  the log-normal
period distribution was an  adequate model.  However, its wings extend
to unrealistically short and long  periods; for this reason we use the
truncated  Gaussian  here.  A  better  analytic  model  of the  period
distribution could be a triangular function with a rounded top,
\begin{equation}
f_T(x) = \frac{1}{Cw}  \left[ 1 - \sqrt{ a^2 +  ((x-x_0)/w)^2 (1 - a^2)}
  \right], 
\label{eq:tri}
\end{equation}
where $|x-x_0| < w$.  This distribution extends over finite range from
$x_0-w$ to $x_0+w$, the parameter $w$ is close to the half-width.  The
parameter $a$ provides for the  rounded top ($a=0$ means pure triangle
$1 -  |x/w|$), the normalization factor $C$  depends on $a$: $C  = 1 -
a^2 (1-a^2)^{-1/2}  \ln [(1 + (1-a^2)^{1/2})/a]$.  The  need for three
parameters $x_0,  w, a$  instead of  just two $x_0,  \sigma$ may  be a
disadvantage  of this  model, but  on  the other  hand the  triangular
distribution does not require truncation.

The ML formalism was adapted to use the triangular function instead of
a Gaussian. The results obtained with this alternative model are quite
similar, they are not given here for brevity. The parameter $a \approx
0.25$ was found, meaning that the  top of the triangle is rounded only
mildly.    The    ``shoulders''   of   the    period   histograms   in
Figure~\ref{fig:phist} are indeed  closer to straight lines  than to a
Gaussian.

Usually the distributions of binary parameters were studied regardless
of hierarchy, merging all  pairs together (DM91, R10).  However, there
is a significant (factor 2.5)  difference between the median period of
{\em all} binaries and the median period of outer binaries at level L1.

The periods  of L11  sub-systems are shorter  than the periods  of all
binaries,  being restricted  by the  dynamical stability.   The period
histogram at L11 deviates from the Gaussian model, showing a depletion
at $x  \sim 2$ and an  excess at $x <1$.   Considering the uncertainty
associated with the missing  data, we cannot evaluate the significance
of  this mismatch.  It is  expected due  to tidal  migration  of inner
sub-systems towards short periods (\S~\ref{sec:Kozai}).

Note  that the  secondary sub-systems  at L12  are only  slightly less
frequent  that at  L11.   The  common belief  that  inner systems  are
preferentially found  around primary components turns out  to be true,
but this  preference is  not strong.  The  median mass of  the primary
components   at   levels   L11   and  L12   is   1.20\,$M_\odot$   and
0.79\,$M_\odot$, respectively.

The  sub-systems at  L12 show  a tendency  towards  equal components,
$\beta =  1.3$.  Only 5 secondary  sub-systems out of  95 have unknown
periods, making it difficult to  explain high $\beta$ by the influence
of the  missing data.  Considering the  bias towards large  $q$ in the
detection of secondary sub-systems (see Figure~\ref{fig:pq}), the high
$\beta$ derived here for L12 should be verified by further work before
being  accepted.   However,  $\beta  \sim  0$  would  mean  that  more
secondary sub-systems  are presently missed, so  their frequency would
be even higher than the frequency of L11.

\section{Statistics of hierarchical systems}
\label{sec:trip}

In  this Section, we  study statistical  relations between  binaries at
different  hierarchical  levels.  Are  their  periods  or mass  ratios
related? Does the presence of one hierarchical level correlate with
other levels? We propose a  statistical model and test it by comparing
simulations with the real sample. 

\subsection{Period ratio and dynamical stability}

\begin{figure}[ht]
\plotone{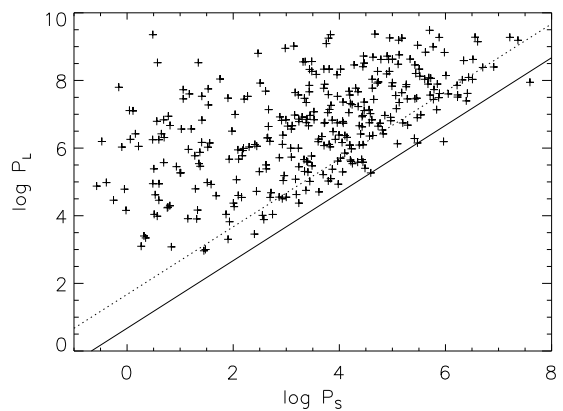}
\caption{Orbital  periods $P_S$ (short)  at inner  hierarchical levels
  L11  and L12  are compared  to the  periods of  outer  systems $P_L$
  (long) at L1.  The periods are  expressed in days and plotted on the
  logarithmic scale.  The solid and dotted lines mark $P_L/P_S$ of 4.7
  and 47, respectively (zone affected by the dynamical stability).
\label{fig:plps} }
\end{figure}

Figure~\ref{fig:plps} compares orbital periods  $P_S$ and $P_L$ at the
inner and  outer hierarchical levels respectively  (only known periods
are  plotted). It  shows that  the  lowest period  ratio $P_L/P_S$  is
limited by the dynamical stability, but this ratio can also take large
values; there  is no  typical or preferred  period ratio.   The points
occupy almost all  the space above the stability  cutoff, showing that
all allowed combinations of periods actually happen.

One striking feature of  Figure~\ref{fig:plps} is the absence of outer
periods  shorter than  $10^3$\,d, as  noted already  by \citet{Tok06}.
This  is  not  caused  by  the observational  selection,  as  tertiary
companions  with  short  periods  are  readily  discovered  by  radial
velocity variation;  the detection probability  at $x<3$ is  high, see
Figure~9 of Paper~I.  \citet{Gies2012} independently discovered the lack
of  short outer periods  by eclipse  timing of  41 binaries  from {\em
  Kepler}: 14  of them show trends indicative  of tertiary companions,
but  none  has  tertiaries  with   $P_L  <  700$\,d.   In  the  3-tier
hierarchies  L11+L111,  the  shortest   period  at  L11  is  $x=2.88$,
confirming the  exclusion zone of  outer periods found for  the L1+L11
and L1+L12  hierarchies.  This said,  triple systems with  short outer
periods are  known among more massive stars  (e.g.  $\lambda$~Tau with
$P_L = 30$\,d).

The  minimum  period  (or   separation)  ratio  allowed  by  dynamical
stability  has been  studied  by several  authors.   For example,  the
stability criterion  of \citet{Mardling}  for co-planar orbits  can be
written as
\begin{equation}
P_L/P_S > 4.7 (1 - e_L)^{-1.8}  (1 + e_L)^{0.6}  (1 + q_{\rm  out})^{0.1} ,
\label{eq:Mardling}
\end{equation}
where $e_L$ is the eccentricity of the outer orbit, while the ratio of
the distant-companion  mass to the  combined mass of the  inner binary
$q_{\rm  out}$   plays  only  a   minor  role.   The  solid   line  in
Figure~\ref{fig:plps}  corresponds to $P_L/P_S  =4.7$; all  points are
above  it (with  two exceptions  caused by  the uncertainty  of period
estimates  from  projected  separations).  \citet{Tok04}  studied  the
dependence of the period ratio on $e_L$ for triple stars with reliable
outer orbits and suggested an empirical stability limit with a
stronger dependence on $e_L$, 
\begin{equation}
P_L/P_S > 5 (1 - e_L)^{-3} . 
\label{eq:Tok04}
\end{equation}

\begin{figure}
\plotone{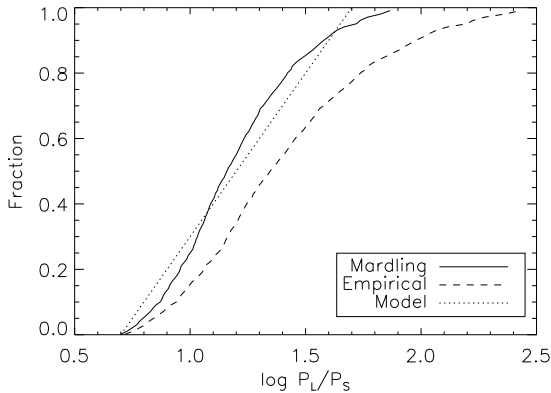}
\caption{Cumulative  distributions of  the dynamical  cutoff $P_L/P_S$
  according to the  equations (\ref{eq:Mardling}) and (\ref{eq:Tok04})
  are  compared  to the  adopted  model  of  the dynamical  truncation
  function $T$ (dotted line).
\label{fig:dynam} }
\end{figure}

At $e_L=0.67$, the limiting period ratio (\ref{eq:Mardling}) becomes a
factor  of 10  larger than  at  $e_L=0$. The  dynamical truncation  of
$P_L/P_S$ is random and extends over at least one decade, depending on
the  distribution of  $e_L$.  Figure~\ref{fig:dynam}  shows the cumulative
distributions  of this  cutoff $\Delta  x  = \log_{10}  P_L/P_S$ if  a
cosine  distribution of  $e_L$  between  0 and  0.8  is assumed,  with
average $e_L  = 0.4$. Available  data on multiple stars  indicate that
orbits  of   outer  systems  tend  to   have  moderate  eccentricities
\citep{Shatsky01}, the  linear distribution $f(e_L) = 2  e_L$ seems to
be excluded.

Here  we use  a  crude  model for  the  dynamical truncation  function
$T(\Delta  x)$  (probability of  a  triple  system  to be  dynamically
stable) by setting  $T=0$ for $\Delta x < 0.7$, $T=1$  for $\Delta x >
1.7$, and  linear in-between (dotted  line in Figure~\ref{fig:dynam}).
This model agrees with the  stability criterion of Mardling \& Aarseth
for the  assumed distribution of $e_L$, while  the  criterion
(\ref{eq:Tok04}) predicts a stronger  cutoff.  An attempt to determine
the   truncation   empirically   by   the  distance   of   points   in
Figure~\ref{fig:plps}  above  the  solid  line  did  not  produce  any
convincing results.  Most outer  and some inner periods are determined
from projected separations  to within a factor of  3, so the empirical
values of  $\Delta x$ are  approximate.  For the same  reason, precise
knowledge of the truncation function  $T(\Delta x)$ is not needed here
and its crude model is sufficient.


\subsection{Independent multiplicity?}
\label{sec:indep}

The  distributions of  both inner  and  outer periods  in the  allowed
region  of Figure~\ref{fig:plps}  look  similar, as  though they  were
drawn from  the same population. Are the  periods at levels L11  and L12 
shorter than at L1 simply because of the dynamical cutoff?

\begin{figure}
\plotone{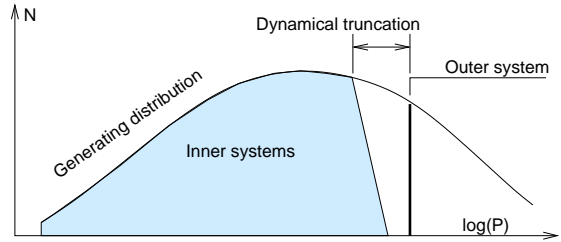}
\caption{Explanation  of independent  multiplicity. The  period  of the
  outer    binary   is    drawn   from    the  log-normal   generating
  distribution. Then  the period of  the sub-system is drawn  from the
  left  part  of  the   same  distribution  fulfilling  the  stability
  constraint (the hatched part)
\label{fig:indep} }
\end{figure}

The idea that the inner and outer pairs of multiple stars are selected
from the same  (or similar) parent distributions, subject  only to the
stability constraint, merits  further investigation. Several arguments
in its favor are furnished by prior work. \citet{Tok06} noted that the
mass  ratios  of close  binaries  with  and  without outer  (tertiary)
companions  are distributed  equally,  although the  former group  has
statistically  shorter   periods.   The  frequency   of  spectroscopic
sub-systems  in  visual  binaries  is  similar  to  the  frequency  of
spectroscopic binaries  in the  open clusters and  field \citep{TS02}.
The frequency and mass ratio  of resolved sub-systems in wide binaries
are again comparable  to the binaries of such  separations in the field
\citep{THH10}.  Finally, the angular momentum vectors in the inner and
outer systems show only  a weak correlation \citep{ST02}, and there
are well documented cases of counter-rotating multiples.

Independent multiplicity is expressed mathematically by setting the
joint period distribution at levels L1 and L11 to be a product of the
individual distributions at these levels, including the  dynamical
truncation factor $T(\Delta x)$:
\begin{equation}
f (x_{L11}, x_{L1}) = f_{L11}(x_{L11} ) f_{L1}(x_{L1}) T(x_{L1} - x_{L11}) .
\label{eq:T}
\end{equation}
Figure~\ref{fig:indep} illustrates  this statistical model.   The {\em
  generating distribution}  for L11, $f_{L11}(x_{L11}  )$, could be
the same as the period  distribution at L1, $f_{L1}(x_{L1})$.  This 
logic  can  be  applied  to   other  levels,  leading  to  a  complete
statistical description of  hierarchical multiplicity. Comparison with
the real data presented below  shows that this model is too simplistic
and needs several adjustments.  However,  to the first order, it works
even in its simplest form.

\begin{figure}
\plotone{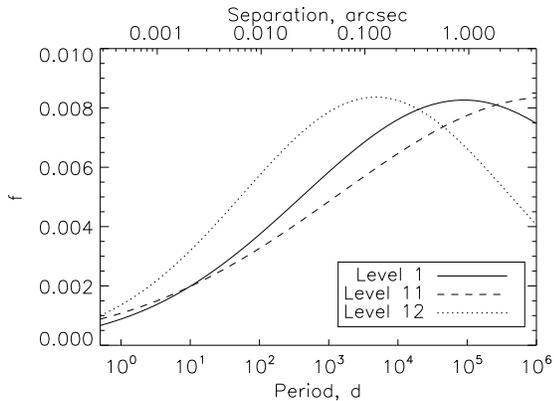}
\caption{Comparison  of  the  generating  Gaussian  distributions  for
  levels L11 and L12 derived by ML with the period distribution of the
  outer systems at L1.
\label{fig:gen} }
\end{figure}

An  attempt was  made to  test  this idea  using the  ML method.   The
truncation  function was  added as  a multiplier  to  the distribution
(\ref{eq:Gauss}) and  the 4 parameters were found  for the sub-systems
at L11 and  L12. The resulting distributions can be  thought of as the
generating  functions $f_{L11}$  and  $f_{L12}$. The  problem of  this
approach   is   that   almost   all  sub-systems   have   $x<6$   (see
Figure~\ref{fig:plps}), so  the data  do not constrain  the generating
functions at longer periods.  This means that the fitted parameters of
the generating  distributions $\epsilon,  x_0, \sigma$ do  not inspire
confidence, and  only the distributions  at $x<6$ (left halves  of the
Gaussians)   are  more   or  less   defined.   Moreover,   a  Gaussian
distribution  is  not  a  good   model  for  the  sub-systems  at  L11
(Figure~\ref{fig:phist}). We  do not  list here the  fitted parameters
and simply plot the relevant  parts of the generating distributions in
Figure~\ref{fig:gen}.  The  periods of L12 sub-systems turn  out to be
shorter and their frequency at $x<4$ is even higher than the frequency
of levels L1 and  L11. However, considering the uncertainties involved
in this crude  model, we can only state that all  three curves are not
very different from each other and that selecting periods at outer and
inner levels from the {\em same distribution} is not such a bad idea.

\subsection{Simulations}
\label{sec:sim}

In this sub-section, we simulate the multiplicity of the FG-67 sample.
A  synthetic  population of  single,  binary,  and  multiple stars  is
created   by  a   random   draw  from   the  generating   distribution
(\ref{eq:Gauss}) with  parameters $(x_0,  \sigma, \beta) =  (5.0, 2.3,
0)$. The period distribution is truncated at $x<-0.3$ and $x>10$.  The
binary frequency $\epsilon$ is specified below.  For each component of
a  binary pair with  $x>3$, the  sub-systems are  drawn from  the same
distribution and  retained with a  probability $T( \Delta  x)$ defined
above.   The  recipe  is  applied recursively  to  produce  high-order
hierarchies.  After that, the  synthetic sample is ``filtered'' by the
average detection  probability of the  real sample (different  for the
primary and secondary components) and compared to the real sample.

Multiple  systems  with  more  than two  hierarchical  levels  (3-tier
hierarchy) pose a specific  problem for their detection.  Suppose that
in   a   quadruple  (((Aa,Ab),B),C)   the   component   B  is   missed
(un-detected).  The system will  be perceived as a triple ((Aa,Ab),C),
i.e.  we miss  the intermediate hierarchical level; the  pair Aa,Ab is
attributed to  L11,  instead of L111.   Now suppose that  the true
system is ((A,(Ba,Bb)),C)  and we miss the same  component B.  In such
case, we can't discover (Ba,Bb),  the observed system becomes a simple
binary  (A,C).  Our  detection  simulator eliminates  all siblings  of
un-detected secondary  components but keeps those of  the primary, and
corrects the hierarchical levels  to reflect the observed, rather than
true, structure of each system.

Straightforward  application   of  the  above  recipe   with  a  fixed
$\epsilon$  over-produces binaries  and under-produces  triples, while
the total number of L1 systems (the multiplicity fraction) is correct.
Therefore, sub-systems  at different hierarchical levels  do have some
correlation.  The simulator is  modified by adopting a variable binary
frequency $\epsilon$ (however, $\epsilon$  is same for each individual
multiple  system).    When  $\epsilon$  is   high,  many  hierarchical
multiples are produced, and when it is low, we get mostly single stars
and simple binaries.  This reflects  the idea that the field sample is
a  mixture of  ``binary-rich'' and  ``binary-poor''  populations.  The
{\it average}  $\epsilon$ is constrained by  the observed multiplicity
fraction $f_M$,  while the relative proportion  of hierarchies informs
us on the  variation of $\epsilon$.  The frequency  of binary stars is
proportional to $\epsilon$, the  frequency of triples to $\epsilon^2$,
etc. By  adopting a  variable $\epsilon$, we  enhance the  fraction of
hierarchies until it matches the real sample (the same result could be
obtained by increasing $\epsilon$ for inner hierarchical levels).

We split the  simulated sample in three equal parts  and apply to each
group  its  own  frequency  $\epsilon_i$.   The  correct  fraction  of
hierarchies was obtained, for  example, for $\epsilon_i = [0.05, 0.60,
  0.75]$.  These  numbers were  found by trial  and error,  their mean
0.466  is  the  simulated  multiplicity  fraction.   Compared  to  the
constant  multiplicity,  the  fraction  of  triples  is  increased  by
$\langle  \epsilon^2 \rangle  /  \langle \epsilon\rangle  ^2 =  1.41$.
There are  other ways  to model variable  $\epsilon$ by  adopting some
distribution  and playing  with its  parameters.  The  distribution of
$\epsilon$ chosen here is discrete with three equally probable values.
Although the values of $\epsilon$ quoted above produce a good match to
the data, other combinations or recipes are not excluded.

The  simulator needs  yet another  adjustment to  reproduce  the large
observed number of 2+2  systems.  If primary and secondary sub-systems
at levels L11 and L12 are generated independently of each other, their
common  occurrence  is too  rare.   The  observed  numbers of  systems
containing  L11,  L12,  and   both  levels  (i.e.   2+2  systems)  are
$[N_{L11},  N_{L12},  N_{2+2}] =  [296,95,41]$.   We  define the  {\it
  correlation}  $C =  N_{2+2}/N_{L12} =0.43$  as the  fraction  of 2+2
systems among the  L12 systems.  Let $N_{L1} = 1747$  be the number of
L1 (outer) binaries in the sample.  If the occurrence and detection of
L11 and L12 were  mutually independent, their respective probabilities
would be  $N_{L11}/N_{L1} = 0.169$ and $N_{L12}/N_{L1}  = 0.054$.  The
product gives an estimate of 2+2  hierarchies as $N_{2+2} = 16$, to be
compared  to the  actual number  $N_{2+2} =  41$.  The  excess  of 2+2
systems indicates correlation between levels L11 and L12.  This effect
was noted by \citet{TS02}:  the frequency of spectroscopic sub-systems
in the  distant tertiary  components of known  triples is  enhanced in
comparison  to the field  stars, as  though the  multiplicity syndrome
were contagious.  Sub-systems of  levels L11 and L12 correlate because
they were formed close to each  other, in the same environment or even
by the same process.

\begin{table}
\center
\caption{Multiplicity counts in real and simulated samples }
\label{tab:comp}
\medskip
\begin{tabular} {l r r |  r r r }
\hline
\hline
Parameter & \multicolumn{2}{c|}{Real} &  \multicolumn{3}{c}{Simulated} \\
          & $N$ & $\langle x \rangle $ &  $N_{\rm obs}$ & $\langle x\rangle $ & $N_{\rm tot}$ \\
\hline
L1  &   1747 & 4.89 & 1776 & 4.91 & 2190 \\
L11 &   296 & 3.4   & 295  & 3.4  & 478  \\
L12 &   95  & 3.7   &  95  & 4.2  & 348  \\
L11 \& L12& 41 &    &  43  &      & 196 \\
L111&   17  & 2.1   & 23   & 2.5  & 49 \\
L112&    4  &       & 5    &      & 33  \\
L121&    3  &       & 12   &      & 36 \\
Single   & 3053  &       &  3024&      & 2610 \\
Binaries & 1397  &       &  1428&      & 1560 \\
Triples  &  290  &       &  274 &      & 366 \\
Quadruples & 55  &       &  62  &     & 204 \\
2+2 quadruples & 37  &   & 35   &     & 151  \\
Quintuples & 5  &        &  9  &      & 43 \\
Hierarchies & 350 &      & 348  &     & 630\\ 
\hline
\end{tabular}
\end{table}

The  excess of  2+2 systems  is accounted  for in  the  simulations by
increasing  $\epsilon$  by $\epsilon_+=1.2$  times  for the  secondary
sub-systems  at  L12, if  the  primary  sub-system  at L11  is  already
present.  Otherwise, the frequency of L12 sub-systems is multiplied by
$\epsilon_-=0.5$; without such  reduction of $\epsilon$, the simulator
over-produces the  level L12.  By playing with  these two  numbers, we
adjust the total number of L12 systems and their correlation with L11.
We  also adopt $\beta=1.0$  for L12,  guided by  the ML  results (when
$\beta \sim  0$ is used for  the secondary sub-systems,  there are more
undetected L12  pairs and the number  of hierarchies becomes
larger, $f_H=0.15$).   We tried several other methods  to simulate the
observed correlation between L12 and L11, for example by modifying the
periods at L12  or by generating L12 with  $\epsilon=1$ in presence of
L11.  The  method chosen here gives  good results, but it is not unique.
Simulations give  the observed correlation $C=0.45$  and the intrinsic
correlation (before applying the detection filter) of 0.56.

A reasonably  good match between  the real and simulated  samples is
demonstrated by Table~\ref{tab:comp}. To reduce statistical errors, we
simulated the sample 10$\times$ larger, $N=48000$, and scaled down the
numbers.   The  simulated  multiples  successfully  mimic  the  period
distributions    of    the   real    sample    and    the   plot  in
Figure~\ref{fig:plps}.  The median periods  at levels L1, L11, and L12
in the simulated and real samples agree well.

The last column of Table~\ref{tab:comp}  gives the system count in the
simulated sample before applying  the detection filter. They represent
the  {\em true}  multiplicity corrected  for  observational selection.
Remember,  however, that  this  method of  correction  depends on  the
simulation  recipe  used here,  which  is  not  unique.  Although  the
simulated and  real statistics  match quite well,  this does  not mean
that  another equally  good simulators  leading to  somewhat different
predictions cannot be devised.   Interestingly, the apparent number of
pure   binaries  remains  almost   unaffected  by   the  observational
selection, while  the numbers of true and  observed hierarchies differ
substantially, especially at level  L12.  The simulations predict that
a large number  ($\sim$150) of 2+2 systems in  the FG-67 sample remain
to be discovered.

The  simulator  over-estimates  the  number  of  sub-systems  at  high
hierarchical levels L111, L121, L112, etc. because it does not account
for tidal evolution in close  inner binaries and selects their periods
from the same  log-normal distribution.  We do not  attempt to correct
this effect and  simply take it into consideration.   Because of this,
the  simulated  number  of  triples  is smaller  than  observed  (274
vs. 290), while the numbers of high-order hierarchies are larger.  The
companion  frequency $f_C$ is  slightly over-estimated.   However, the
total number  of hierarchies, which depends  only on levels  L11 and L12
and on their correlation, is simulated correctly.

The total  number of hierarchical  systems in the simulated  sample is
630, leading to $f_H = 0.132$.  It agrees with $f_H = 0.129$ estimated
by  the ML method  (different experimental  versions of  the simulator
produced $f_H$ between 0.12 and 0.15).  The fractions of L11, L12, and
2+2  systems in  the simulated  sample  are 0.100,  0.073, and  0.041,
respectively.  The estimated fractions of single, binary, etc. systems
(last column of Table~\ref{tab:comp})  are $f_n = [0.543, 0.323, 0.076,
  0.043, 0.013]$,  where the  last number is  the fraction  of systems
with 5 or more components.   The calculated companion fraction is then
$f_C  =  0.65$.   As expected,  it  is  larger  than its  ML  estimate
$\epsilon =0.57$  (the first line in  Table~\ref{tab:par}). The actual
companion fraction must be about 0.6.

It  is truly remarkable  that   complex statistics  of hierarchical
multiple systems  can be reproduced  by the simulation recipe  with only
five adjusted parameters $\epsilon_i$, $\epsilon_+$, and $\epsilon_-$,
while the  remaining parameters  of the generating  distribution $x_0,
\sigma,  \beta$ are determined  by the  ML method  and kept  fixed. This
supports the  underlying idea of  independent multiplicity. Therefore,
the shorter periods of sub-systems  at levels L11 and L12 (compared to
all binaries) can be explained by the dynamical truncation alone.

\subsection{Tidal evolution}
\label{sec:Kozai}

\begin{figure}
\plotone{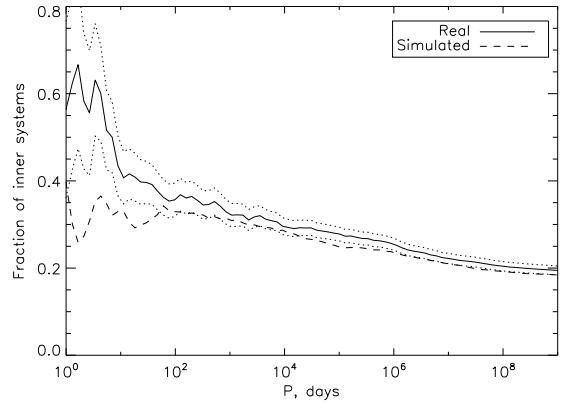}
\caption{The  observed  fraction of  inner  sub-systems (levels  $>$L1)
  among all  pairs up to a given  period is plotted in  the full line.
  The  dotted  lines show  the  statistical  errors,  the dashed  line
  corresponds to the simulated sample.
\label{fig:multfrac} }
\end{figure}

Figure~\ref{fig:multfrac} illustrates  one aspect of the  data that is
not  yet captured  by the  simulator.  All  binaries up  to  a certain
period $P$ are  selected, and the observed fraction  of inner pairs in
multiples  (levels higher than  L1) among  those binaries  is plotted.
The  fraction of  inner  systems reaches  $\sim 0.6$  at $P<3$\,d  and
slowly drops  to 0.18 with  increasing period.  The fraction  of inner
pairs  in the simulated  sample (with  the detection  filter applied),
plotted in dashed  line, is quite similar to  the observed one, except
at $P < 10$\,d where it is much lower.

Inner  binaries  in  triple  systems  with  mutually  inclined  orbits
experience   periodic  modulation   of  their   eccentricity   by  the
Kozai-Lidov  cycles.   When   the  separation  at  periastron  becomes
comparable to  the stellar radii,  tidal friction absorbs  the orbital
energy and  the period of the  inner binary shortens  to $P_S \la 10$\,d.
This   mechanism,  known   as   Kozai  cycles   with  tidal   friction
\citep[KCTF,][]{Eggleton2006},  works  well   when  the  period  ratio
$P_L/P_S$ is  not too high and  the initial inner period  $P_S$ is not
too long. It causes migration  of inner periods from $P_S \sim 100$\,d
to $P_S  \le 10$\,d \citep{Fabrycky07}, while some  inner binaries can
merge.

Tidal  migration  of   inner  periods  reflects  in  their  period
distribution  (lower panel  in  Figure~\ref{fig:phist}), although  the
addition  of unknown  fictitious  periods to  the histogram  partially
masks the effect.  See also the cumulative period histogram at L111 in
Figure~\ref{fig:pcum},  where   40\%  of  periods   are  shorter  than
$\sim$3\,d.  The period distribution  of spectroscopic binaries in the
Hyades \citep[][figure 68]{Griffin2012} clearly shows the depletion at
$P>10$\,d and the peak  at shorter periods.  The non-monotonous period
distribution was also found by \citet{Halb03}.
Tidal evolution explains why the median period at L111 is shorter
that  its simulated  value and  why the  simulator  over-estimates the
number  of high  hierarchical  levels.  Apparently,  some close  inner
binaries merged.

\subsection{Mass ratios in hierarchical systems}
\label{sec:q}

\begin{figure}
\plotone{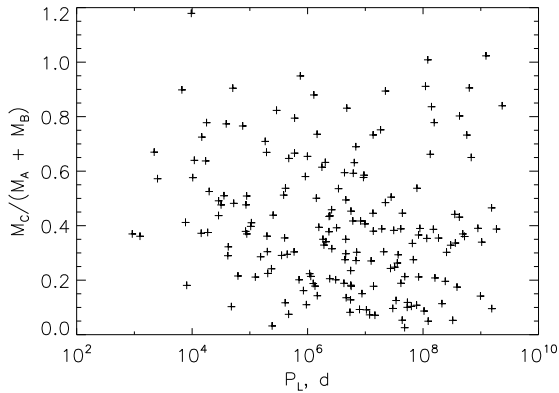}
\caption{System mass  ratio $q_{\rm sys}  =  M_C/(M_A  +  M_B)$ in  the  outer
  (level L1) systems  of triple  stars as a  function of the outer orbital
  period $P_L$.
\label{fig:q3} }
\end{figure}

We   examined  correlations   between  the   mass  ratios   in  triple
systems. The  199 simple  triples (levels L11  or L12  without further
inner sub-systems, with known periods  and masses) are used.  The data
are  affected to  some extent  by the  observational selection,  as no
simple  correction  can  be  suggested  in this  case.   We  found  no
correlation between  the mass  ratios in inner  and outer  systems and
between the mass ratio and the period.  Similar conclusion was reached
by \citet{Tok08}  from analysis of  the biased catalog.   For example,
Figure~\ref{fig:q3}  plots  the ratio  $q_{\rm  sys}$  of the  distant
companion's  mass to the  mass sum  of the  inner binary.   The median
value of  $q_{\rm sys}$  for outer binaries  is 0.39, with  no obvious
dependence on the outer period $P_L$  (when the sample is split in two
halves at $P_L = 10^{6.7}$\,d,  the $q_{\rm sys}$ medians are 0.40 and
0.39).  The median mass of the tertiary components is 0.73\,$M_\odot$,
the median mass  of the secondary components in  {\em all} binaries is
0.70\,$M_\odot$.   Therefore, there  is no  evidence that  the distant
tertiary components are less massive  than the components of the inner
binaries.

\citet{Tok08} noted  that in 2+2  quadruples with $P_L  < 10^{5.4}$\,d
both inner binaries have similar  total masses and periods.  The FG-67
sample  contains only  5  such  quadruples, all  with  $q_{\rm sys, out}  >
0.5$. The statistics are too small to make any further conclusions.

We checked whether  twin binaries have any preference to be found
in multiple systems. Half of  the 96 binaries with $P<10$\,d are inner
sub-systems (see Figure~\ref{fig:multfrac}).  Among those 96 binaries,
18 are  twins with $q>0.95$,  and 8 of  those twins (also a  half) are
in inner sub-systems.  Therefore, the  phenomenon of twin binaries is not
directly related to the hierarchical multiplicity.  The mass ratios of
all  binaries,  including  twins,  seem  to be  independent  of  their
hierarchical level.

One  notable  exception  to  the above  statement  concerns  secondary
sub-systems  at L12  which show  a  tendency to  equal masses,  $\beta
=1.3$,      according       to      the      ML      parameter-fitting
(\S~\ref{sec:param}). About half of those binaries (41 out of 95) also
have primary  sub-systems at  L11. The median  mass ratio in  those 41
primary  sub-systems is  0.75,  while the  remaining  majority of  L11
sub-systems have  median mass ratio of  0.65. 
This  result  does  not  rely  on  the  ML  analysis  or  completeness
correction,  it  is a straightforward  comparison between  the  primary
sub-systems  whose outer  companions are  themselves either  binary or
single.  Preference  of equal  masses is thus  typical for  {\em both}
inner pairs  in  2+2 hierarchies.  This  suggests that  they could
have formed differently, compared to other binary and triple stars.

\section{Summary and caveats}
\label{sec:sum}

\begin{figure}
\plotone{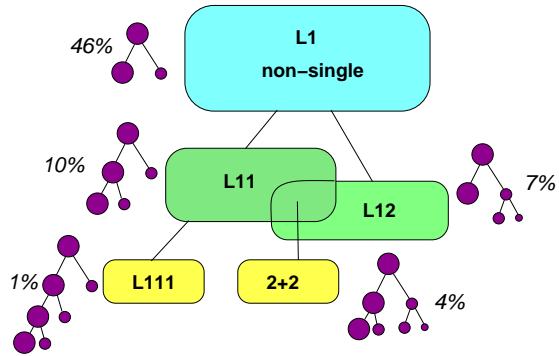}
\caption{Fractions of different hierarchies  in the FG-67 sample.  The
  levels L11 and  L12 overlap by 4\%, accounting  for the 2+2 systems,
  while the total fraction of hierarchies is $10+7-4=13$\%.
\label{fig:stat} }
\end{figure}

The main results of this study  are:

\begin{enumerate}

\item
We confirm  known facts about  binary statistics of  solar-type stars.
Periods  of   all  binaries   (regardless  of  their   hierarchy)  are
distributed log-normally  with a  median $x_0 =  4.54$ ($\sim$100\,yr)
and dispersion  2.4.  The mass  ratio is distributed uniformly  and is
independent (or almost independent) of the period, except such details
as short-period twins. The multiplicity fraction is $0.46 \pm 0.01$.

\item
The fraction  of hierarchical systems is 0.13$\pm$0.01,  of which 0.10
have  sub-system(s)  related  to  the  primary  component,  0.07  have
sub-systems   in  the   secondary  component,   and  0.04   have  both
(Figure~\ref{fig:stat}).  The  presence  of  sub-systems  in  both
components of wide (outer) binaries is correlated.


\item
There  remain  about  150  undiscovered secondary  (L12)  sub-systems.
About 60\%  of those pairs reside  in  2+2  hierarchies, because of
the above correlation.

\item
The lack of outer systems with periods shorter than $\sim$1000\,d is real
(not a selection effect). 

\item
The statistics of hierarchical  systems can be reproduced by recursive
random selection of outer and  inner binaries from the same generating
period  distribution  with  parameters  $x_0=5.0$,  $\sigma=2.3$,  and
uniform   mass  ratio,  $\beta=0$   ($\beta=1.0$  for   the  secondary
sub-systems).  The  amplitude of  this distribution is  variable (e.g.
$\epsilon=[0.05,0.60,0.75]$ with equal probability).  Only dynamically
stable  combinations are  kept,  with the  dynamical  cutoff in  $\log
(P_L/P_S)$ uniformly distributed between 0.7 and 1.7.  The frequency of
secondary  sub-systems  is  enhanced  by their  correlation  with  the
primary sub-systems, while binaries  with periods shorter than 1000\,d
do not produce sub-systems.

\item
The period distribution  in the inner sub-systems can  be explained by
dynamical  truncation.  However,  there is  an excess  of  tight inner
binaries with $P<10$\,d compared  to the smooth Gaussian distribution,
presumably caused by tidal evolution.

\item
The mass  ratios in the  inner and outer  systems of triple  stars are
uncorrelated, except 2+2 hierarchies  where the mass ratios are larger
than average in both of  their inner sub-systems. In triple stars, the
system mass  ratio of the outer  binary does not depend  on its period
and has  a median value of  0.39, meaning that the  masses of tertiary
components are comparable to the masses of stars in the inner binary.

\end{enumerate}

The  above  results are  not  free  from  biases, approximations,  and
errors. Some were mentioned above and in Paper~I. We re-capitulate the
most relevant {\em weaknesses} below, in order of decreasing importance.

\begin{enumerate}

\item
{\em  Missing information.} Unknown  periods and  mass ratios  of many
spectroscopic and astrometric binaries (32\% of all binaries with $P <
100$\,yr) increase the uncertainty of the results. Missing information
is accounted for in the  ML analysis and in the simulations.  However,
we cannot reliably  study the distribution of periods  and mass ratios
of close binaries until they are resolved or get spectroscopic orbits.

\item
{\em  Approximate  knowledge of  detection  probability.} Despite  the
effort to  collect relevant  data and to  model the  detection limits,
there  remains  some  uncertainty.   For example,  all  binaries  with
separation of $\sim1''$  and $\Delta V <4$ are  assumed to be resolved
by {\em Hipparcos},  but this is not actually  the case. The detection
of sub-systems  in the  secondary components is  most critical,  as it
entails a larger correction for incompleteness.

\item
{\em Uncertain data and subjective decisions.}  Some binaries accepted
as real  may in fact be  spurious or optical. Conversely,  we may have
rejected  or missed  some real  pairs (see  \S~3.10 of  Paper~I).  The
percentage  of such  cases is  certainly much  less than  10\%  of all
binaries, but we do not know how much less.

\item
{\em Biases  against multiple  stars.}  Secondary components  that are
themselves binary can  be partially resolved and, for  this reason, be
missed by  the point-source catalogs.  Their proper  motion may differ
substantially  from that  of  the main  target.   The {\em  Hipparcos}
parallaxes of unrecognized binaries  are often biased by their orbital
motion, relegating some of them beyond the 67\,pc distance limit. Such
nearby multiples as $\zeta$~Cnc  and $\xi$~UMa are missed here because
they are not included in the {\em Hipparcos} catalog.

\item
{\em White dwarfs.} The  fraction of Sirius-like binaries harbouring a
WD companion is estimated as $\sim$2\%.  Known Sirius-like systems are
removed  from the  analysis,  but  the reduction  of  the sample  size
associated  with the  removal of  WDs is  uncertain.  Most  likely, the
assumed  sample size  $N_{tot}=4800$  should be  reduced further,  the
estimated $f_M$ will then increase slightly.

\item
{\em Approximations} made in  the data analysis: relation between mass
and  absolute magnitude  in various  photometric bands,  evaluation of
period from projected separations,  evaluation of the mass ratio from
photometry  or from  the  minimum mass  of single-lined  spectroscopic
binaries,  distribution of  periods  and mass  ratios approximated  by
analytical functions.

\end{enumerate}

It is clear  that the results of this study are  affected by  these
biases  and that  the  formal errors  quoted  above underestimate  the
actual,  larger uncertainties.  Future  work will  revise some  of the
parameters derived  here. However,  these anticipated changes  will no
longer be dramatic; the  fraction of hierarchical systems derived here
should not be revised by more than 1-2\%.

\section{Discussion}
\label{sec:disc}

\subsection{Comparison with previous work}
\label{sec:comparison}

The large  sample of solar-type  binaries with known  detection limits
confirms   and  strengthens   known  facts   about   their  statistics
\citep{DK13}.   The parameters of  the log-normal  period distribution
determined here  ($x_0=4.54, \sigma=2.40$)  match well the  results of
R10   ($x_0=   5.03$,   $\sigma=2.28$).    The   triangular   function
(\ref{eq:tri}) fits the distribution of  $x = \log P$ even better than
a Gaussian. However, the  period distribution of the inner sub-systems
is  not log-normal,  it has  an excess  at $P<10$\,d.

We find  a nearly flat distribution  of the mass ratio  with the power
index $\beta \sim  0$ and show that it does not  depend on the orbital
period   (with  a   caveat   related  to   partially  missing   binary
parameters). \citet{DK13} fitted  $\beta = 0.28 \pm 0.05$  to the data
of R10, but found that $\beta = 1.16 \pm 0.16$ for periods shorter than
$10^5$\,d.   Considering the  sensitivity  of derived  $\beta$ to  the
treatment of  selection effects (see  \S~\ref{sec:beta-p}), we believe
that this claim  is premature.  \citet{RM13} derive $\beta  = 0.25 \pm
0.29$ for the two combined samples of dwarf binaries and argue for the
independence of $\beta$  on period and primary mass.   The flat $f(q)$
was  found by  \citet{Kraus2011}  for the  pre-main-sequence stars  in
Taurus more massive than $0.7 M_\odot$. 

This study gives the first look at unbiased statistics of hierarchical
systems with three or more components.  Multiples are frequent (one in
eight  stars).  The 13\%  frequency of  hierarchies estimated  here is
essentially  the  same as  12\%  found  by  \citet{R10} in  the  25-pc
sample. However, they  focused on companions to the  primary stars and
missed some secondary pairs. A  few of those secondary sub-systems are
recovered here,  more are  yet to be  discovered.  On the  other hand,
these  authors  may  have  over-estimated  the  number  of  high-order
multiples.   They list three  multiples with  five or  more components
(their  Fig.~24), while  we  found only  5  such systems  in a  sample
10$\times$  bigger.   This  discrepancy  is explained  by  the  better
companion  census  of nearby  stars,  by  the  inclusion of  uncertain
sub-systems (such as HD~186858 Aa,Ab)  in the 25-pc sample, and by the
small-number statistics.

Note that  among the 11  quadruples in the  25-pc sample, 9  (or 82\%)
have the 2+2 hierarchy and only  two have the 3+1 hierarchy.  The last
column  of  Table~\ref{tab:comp}  predicts  a  74\%  fraction  of  2+2
quadruples, in agreement with the small statistics within 25\,pc.

\citet{Fuhrmann11} gives  estimates of $f_M$ and $f_H$ among F-type
stars within 25\,pc  that agree well with our  results. He claims that
both  fractions rise  sharply with  increasing primary  mass (however,
with a low significance limited by the sample size). We do not confirm
this  trend:  the   observed  (not  selection-corrected)  multiplicity
fraction of 845 stars with  $M > 1.3\,M_\odot$ within 67\,pc is $38\pm
2$\%, not significantly different from 36\% in the full FG-67 sample.

\subsection{Statistical model of multiplicity}
\label{sec:model}

The  statistics   of  hierarchical  multiples  are   well  modeled  by
independent selection of sub-systems from some generating distribution
of periods, provided that the binary frequency is variable. This hints
that  the   field  stars   were  formed  in   different  environments,
binary-rich  and binary-poor.   Indeed, the  binary  frequency differs
between   star-forming  regions   \citep{King2012}.    Moreover,  some
fraction of single stars  come from disintegrated multiples and merged
binaries, so  the {\em intrinsic}  frequency of the  formation process
$\epsilon$ is  always larger than  the multiplicity of a  mature field
population.

Variable multiplicity  naturally leads to the  enhanced probability of
finding sub-systems  in wide binaries,  in comparison to  all targets.
This  tendency  was   noted  by  \citet{Makarov08}.   The  simulations
faithfully  reproduce  the fraction  of  close  binaries belonging  to
higher-order  systems   for  the  inner  periods   longer  than  10\,d
(Figure~\ref{fig:multfrac});  at shorter  inner periods,  the observed
fraction is much higher than predicted, because of tidal evolution.

\citet{Kraus2011} already tried  to model hierarchical multiplicity in
a  Monte-Carlo  simulation by  producing  inner  sub-systems with  the
generating  period  distribution and  applying  the stability  cutoff.
Similar approach  was taken by \citet{Santerne2013}  in their modeling
of  hierarchical  multiples  for  evaluation  of  false  positives  in
exo-planet  detection.  The multiplicity  simulator developed  here is
anchored to the data and can be used in such studies with confidence.

\subsection{The 2+2 systems}
\label{sec:corr}

Correlated presence of sub-systems in both components of a wide binary
is the new fact found in the course of this project \citep{RoboAO}. To
reproduce this  correlation by simulation, the  frequency of secondary
sub-systems  is enhanced  in presence  of the  primary  sub-system and
decreased otherwise.   As a result, the majority  of quadruple systems
have the 2+2 hierarchy (67\% observed, 74\% predicted after correction
of observational  selection, 82\% in the 25-pc  sample).  The fraction
of quadruple  stars is then  larger than predicted by  the independent
multiplicity, and comparable to the fraction of triples (4\%
and  8\%,  respectively).  The  multiplicity  fractions  $f_n$ do  not
decrease in  geometric progression of  $n$, as suggested in  the early
works \citep{Batten} and repeated  by \citet{DK13}.  Poor detection of
sub-systems in  secondary components  was not fully  appreciated until
recently.

The relatively large  fraction of 2+2 systems suggests  that they were
formed  by some special  process. Other  properties of  2+2 quadruples
such as  similarity of component's  masses and inner periods  point in
this  direction, too  \citep{Tok08}.  Both  inner sub-systems  in the 2+2
hierarchies differ from the rest  of inner sub-systems by their larger
mass ratios (\S~\ref{sec:q}).  On the  other hand, when only {\em one}
inner sub-system is present,  it is preferentially associated with the
most massive star, and its  mass ratio is distributed uniformly, as in
normal binaries.  Therefore, there must be another, dominant formation
channel which produces most binary and triple stars and, by extension,
3+1 quadruples.

\subsection{Implications for multiple-star formation}
\label{sec:form}

Statistical work on  multiple stars, including this one,  is motivated by
the desire  to understand their  formation. What have we  learned from
this study?

Hydrodynamical simulations  \citep{Bate2012} reveal a  complex process
of binary  formation and early evolution that  shapes the multiplicity
statistics.   Stellar   pairs  form  by   fragmentation  with  initial
separations larger than 10-100\,AU (so-called opacity limit) and small
masses. The  components grow by accreting  gas and, at  the same time,
migrate to  smaller separations.  The  depletion of long  periods (the
right  side of the  ``Gaussian'') by  dynamical interactions  with the
environment,  combined  with  migration,  shape  the  observed  period
distribution \citep{Korntreff12}.   If the fragmentation generates periods
$x = \log P$ longer than, say,  $x=4$ (a step function in $x$) and the
period  decay  $\Delta  x$  is  a  random  function  (e.g.   uniformly
distributed), the  convolution of  those two distributions  results in
the linear decrease of binary frequency at $x<4$, as actually observed
(Figure~\ref{fig:phist}).   The approximately  linear decrease  of the
period  distribution  at $x<4$  is  produced  under  a wide  range  of
assumptions, provided only that  the initial period distribution has a
lower cutoff and that the migration is random.

Migration  is   essential  in   understanding  the  origin   of  tight
hierarchical systems where 3 or 4  stars are packed in a small volume.
If  all inner  binaries  formed  with initial  periods  $x \ge4$,  the
initial  outer periods must  be even  longer, whereas  the present-day
outer  periods  can  be  as short  as  $x=3$  (Figure~\ref{fig:plps}).
Migration  occurs  therefore at  all  hierarchical  levels. The  inner
binaries must migrate faster or  form earlier than the outer binaries,
otherwise the multiple system becomes dynamically unstable and decays.

Most hierarchies seem to be assembled ``from inside-out'', by bringing
together  pre-fabricated inner  pairs  (or stars)  to  make the  outer
system.   The   success  of   modeling  the  observed   statistics  by
independent multiplicity strongly supports this view.  The mass ratios
of  inner and  outer systems  do not  correlate, the  masses  of outer
(tertiary)  components  are comparable  to  the  masses  of the  inner
components.  The  inner and  outer periods are  selected independently
from  the same  distribution  and  are mutually  related  only by  the
dynamical   stability  (taking   aside  the   tidal   evolution).   If
hierarchies formed ``from  outside-in'' by successive fragmentation of
outer binaries, as envisioned  e.g.  by \citet{Kraus2011}, it would be
difficult to explain the independent multiplicity.

Considering that  many hierarchies are  close to the  stability limit,
dynamical  decay of  multiple systems  must occur  sometimes.   It can
happen when  the outer binary migrates faster  than the inner  one or
when  other stars  in the  cluster  disrupt the  multiple or  exchange
components with  it.  However,  multiple systems produced  by $N$-body
dynamical interactions  (or surviving them)  have characteristics that
do  not match  real hierarchies:  high eccentricities,  low  masses of
distant  components,  and  moderate  ratios $P_L/P_S$.   The  observed
multiplicity statistics indicates that dynamical processes play only a
secondary role. Most hierarchical stellar systems in the field are not
the surviving remnants of primordial clusters.

The  2+2  hierarchies could  be  formed in  a  special  way, e.g.   by
inelastic collision of two cores that prompts their fragmentation into
sub-systems and, at the same time, keeps these two pairs together on a
common    wide    orbit    by    dissipating   the    energy.     Such
impulsively-triggered   multiple-star  formation  was   envisioned  by
\citet{Whitworth}, see his Figure~2.  Further evolution will depend on
the relative orientation  of the inner and outer  orbits.  If they are
not aligned,  the inner  binaries will accrete  gas with  a misaligned
angular  momentum,  which will  accelerate  their  migration to  short
periods.   In the opposite  case of  co-aligned orbits,  the accretion
will keep  the inner  binaries wide, and  a quadruple with  a moderate
$P_L/P_S$    ratio   and    nearly   co-planar    orbits,   resembling
$\varepsilon$~Lyr,  can be  formed.  Another  striking aspect  of this
prototypical 2+2 quadruple -- similar masses of all 4 stars -- matches
the tendency  to equal component  masses in both inner  sub-systems of
2+2 hierarchies, emerging from this study.

On the other  hand, the majority of the  inner sub-systems have $\beta
\sim  0$,  like all  binaries,  while  their  tertiary components  are
single.  Their  formation process  should therefore be  different from
the formation  of 2+2 hierarchies.  The tertiary  component could have
formed around the initial binary (or joined it) at a later stage.  The
cascade  can continue  further by  adding yet  another star  in  a 3+1
hierarchy.

The  2+2  formation channel  probably  produced  more quadruples  than
actually observed because some  sub-systems  merged, leaving triples
and  binaries,  while some  quadruples  decayed  dynamically. A  certain
fraction of  binaries and triples  in the field  could be born  as 2+2
systems. Also, merging of the inner binaries (natural extension
of migration) can possibly help in understanding why the outer periods
in triple stars are longer than $10^3$\,d.

It  should be stressed  that not  {\em all}  close binaries  belong to
multiple stars. This is true  only for pairs with $P<3$\,d which could
not  be produced  without tidal  migration during  their main-sequence
life.  At longer periods $P \ge 10$\,d, the frequency of wide tertiary
companions is  comparable to the  frequency of such companions  in the
full sample \citep{Tok06}.

\begin{figure}
\plotone{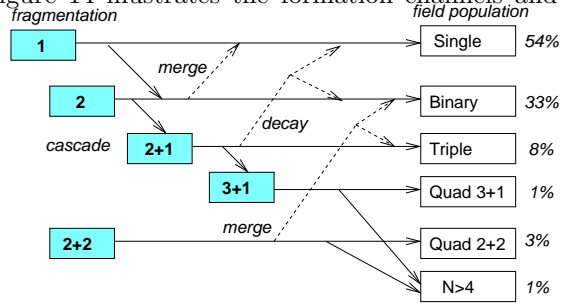}
\caption{Formation  channels  of  multiple  systems and  fractions  of
  different hierarchies in  the field. 
\label{fig:form} }
\end{figure}

Figure~\ref{fig:form}  illustrates  the  formation  channels  and  the
fractions of  different hierarchies  in the field.   The fragmentation
produces   single   and    binary   stars.    Cascade   ``inside-out''
fragmentation  generates  2+1 triples  and  3+1 quadruples.   Unstable
hierarchies decay dynamically into single and binary stars, some close
binaries merge.   The alternative 2+2 process  creates quadruple stars
with components  of similar mass.  Merging  of inner pairs  in the 2+2
quadruples  contributes to  the  populations of  triples and  binaries.
About 1\% of targets with more  than 4 components could be produced by
the combination of these elementary processes.

\subsection{Concluding remarks}
\label{sec:rem}

There are various ways to improve and continue the present study.  The
determination of unknown periods and  mass ratios of close binaries is
one obvious project  to pursue. The prediction of  the large number of
undiscovered sub-systems in the  secondary components should be tested
observationally.  It will be  interesting to study relative motions in
resolved  triple and quadruple  systems to  get a  better idea  of the
relative orbit orientation.



\acknowledgments  The  data used  in  this  work  are acknowledged  in
Paper~I.  I thank Bo~Reipurth and M.~Bate for comments on the draft of
this paper, P.~Eggleton for discussions of multiplicity statistics and
white dwarfs.

\appendix

\section{Maximum likelihood formalism}
\label{sec:ML}

We want to model the data -- periods and mass ratios of binaries -- by
an    analytical    distribution    function    $f(x,q)$    such    as
equation~\ref{eq:Gauss},  i.e.  find  the  best-fitting parameters  of
this model  and their errors.  Some data are missing  (unknown periods
and mass ratios).

Let the  sample of $N$ targets  contain $K$ known  binary systems with
period  logarithms  $x_k$ and  mass  ratios  $q_k$.   For each  target
(binary  or   single)  we   also  estimate  the   companion  detection
probability  $d_i(x,q)$.  We  maximize the  {\it  likelihood function}
${\cal L}$  -- the probability of  getting the actual  data, given the
model parameters.  The maximum  likelihood (ML) approach used here is
equivalent to the Bayesian method \citep{Allen07} with uniform priors.

For  each  $i$th  star,  the  probability  $p_i$  follows  the  Poisson
distribution  $p(m)  =  (\mu/m!)  \exp(- \mu)$  with  parameter  $\mu_i$
(probability of  a detectable companion) and the  integer argument $m=
0$ for single stars and $m=  1$ for binaries. The stars are
observed independently of each other,  so ${\cal L}$ is the product of
$p_i$.   Instead of  maximizing ${\cal  L}$, we  minimize  its natural
logarithm $S = - 2 \ln {\cal L}$. Elementary derivation leads to
\begin{equation}
S = 2 \sum_{i=1}^N \mu_i - 2 \sum_{k=1}^K \ln \mu_k .
\label{eq:S}
\end{equation}
The  summation over $k$  extends only  over detected  companions.  The
first term  equals $N  \mu_0$, $\mu_0$ is  the average  probability of
companion detection.  The probabilities  for the binary companions are
$\mu_k =  f(x_k,q_k) d_k(x_k,q_k)$, while  for all targets  $\mu_i$ in
the first  term are the  products $f(x,q) d_i(x,q)$ averaged  over the
considered  part of  the  parameter space  $(x,q)$. Any  normalization
factor in  the probabilities $\mu_k$ for binaries  that is independent
of the parameters has no influence on the result.

For  binaries with  known periods  but unknown  secondary  masses, the
products $f(x_k,q)  d_i(x_k,q)$ are  averaged over $0  < q  <0.8$. The
rationale is that binaries with  $q>0.8$ would have been recognized as
double-lined  or  resolved,  therefore  the unknown  mass  ratios  are
somehow  constrained. Binaries where  both period  and mass  ratio are
unknown  have the  detection probabilities  averaged over  $q<0.8$ and
also over periods,  assuming $x<4.5$. It is important  to include both
kinds of binaries with partial information in the calculation, as they
are actually discovered.

For  each binary,  the  detection probabilities  are  computed on  the
2-dimensional grid  with 100 points  in $x$, $-0.3  < x < 10$,  and 20
points in $q$.  Calculation of detection probability for the secondary
components is  done by a  different program, as explained  in Paper~I,
and  saved  in   a  different  array  of  the   same  dimension.   The
distribution function $f(x,q)$ is computed on the same grid, the first
bin  in  $q$  is set  to  zero  (no  companions with  $q<0.05$).   The
normalization  constant $C$  in equation~(\ref{eq:Gauss})  is computed
numerically.

The ML formalism  was tested on the simulated  samples. Here the model
and the sample match by  definition, and the ML method indeed recovers
the  true  parameters of  the  distributions,  even  when a  realistic
fraction of binaries is assumed  to have unknown periods and mass ratios,
as in the real sample.

For calculation of  the generating distributions at levels  L11 and L12,
the  ML algorithm  was modified  to include  the  dynamical truncation
function $T(\Delta x)$. For calculating the mass-ratio distribution in
selected intervals of period, the detection probabilities are averaged
over these intervals, leaving only their dependence on $q$.  Then only
two parameters $\epsilon, \beta$ are  fitted to the data.

The contours of  $S$ in the parameter space  define confidence limits:
$\Delta S = 1$ corresponds to the 68\% interval (1$\sigma$), $\Delta S
= 2.71$ to 90\%, etc., in direct analogy with the Gaussian probability
distribution.   The error  of some  parameter $\xi$  is  computed from
the one-dimensional function  $S(\xi)$, while other  parameters are fixed.
This is correct  if the parameters are not  correlated. We checked the
absence  of  strong correlation  by  plotting the contours  of $S$  while
varying any two parameters around  their best values. The contours are
more  or less round,  indicating the  lack of  significant correlation
between the parameters.

The  ML  method  appears  more  rigorous than  it  actually  is.   Its
systematic  errors (biases)  are caused  by the  mismatch  between the
actual data  and their model  (both the parametrized  distribution and
the  detection probabilities).  The  confidence intervals  returned by
the ML  describe only  the statistical errors  and do not  account for
these systematic  errors.  Confidence in  the ML results for  the real
sample is gained by checking robustness with respect to slight changes
in  the models,  alternative probability  calculation,  application to
sub-samples, etc.


\end{document}